\documentclass[aip,jcp]{revtex4-1}  
\usepackage{graphicx}

\usepackage{epstopdf}
\usepackage{natbib}
\usepackage{amsmath} % need for subequations

\begin{document}

\title{Pair Interaction Potentials of Colloids by Extrapolation of Confocal Microscopy Measurements of Collective Suspension Structure }
\author{Christopher R. Iacovella}
\affiliation{Department of Chemical Engineering  \\University of Michigan, Ann Arbor, Michigan 48109-2136}
\author{Reginald E. Rogers}
\affiliation{Department of Chemical Engineering  \\University of Michigan, Ann Arbor, Michigan 48109-2136}
\author{Sharon C. Glotzer}
\affiliation{Department of Chemical Engineering  \\University of Michigan, Ann Arbor, Michigan 48109-2136}
\affiliation{Department of Materials Science and Engineering \\University of Michigan, Ann Arbor, Michigan 48109-2136}
\author{Michael J. Solomon}
\affiliation{Department of Chemical Engineering  \\University of Michigan, Ann Arbor, Michigan 48109-2136}

\date{\today}

\begin{abstract}
A method for measuring the pair interaction potential between colloidal particles by extrapolation measurement of collective structure to infinite dilution is presented and explored using simulation and experiment. The method is particularly well suited to systems in which the colloid is fluorescent and refractive index matched with the solvent.  The method involves characterizing the potential of mean force between colloidal particles in suspension by measurement of the radial distribution function using 3D direct visualization. The potentials of mean force are extrapolated to infinite dilution to yield an estimate of the pair interaction potential, $U(r)$.  We use Monte Carlo (MC) simulation to test and establish our methodology as well as to explore the effects of polydispersity on the accuracy. We use poly-12-hydroxystearic acid-stabilized poly(methyl methacrylate) (PHSA-PMMA) particles dispersed in the solvent dioctyl phthalate (DOP) to test the method and assess its accuracy for three different repulsive systems for which the range has been manipulated by addition of electrolyte. 
\end{abstract}

\maketitle

\section{Introduction}

Colloidal systems may undergo crystallization because of the effects of packing and excluded volume \cite{pusey1987}, repulsive or attractive charge \cite{sirota1989, leunissen2005}, and/or weak attractions caused by additives such as non-absorbing polymer \cite{ilett1995}. Moreover, the phase behavior can often be dramatically changed by making only small modifications to the particle-particle interactions \cite{hagen1994, azhar2000, leunissen2005}.    Knowledge of the interaction between colloidal particles and the ability to tune these interactions is important for designing and assembling target materials. Target materials, such as self-assembled arrays of colloids, have potential applications as periodic dielectrics \cite{hosein2007}, photonic band gap materials \cite{ngo2006}, and chemical and biological sensors \cite{hosein2007b}. The pair interaction potential, $U(r)$, which characterizes the potential energy change that results as two isolated particles are brought from an infinite to a finite separation, is a common parameterization of the interaction between colloids.  The ability to parameterize particle interactions into pair potentials allows simulation and theory to be readily incorporated into the experimental design process.  Simulation and theory have been shown to be valuable tools for predicting and explaining structures and trends in colloidal systems. For example, simulation and theory have been used to calculate the crystal-nucleation rate of hard-sphere colloids \cite{auer2001}, phase behavior of attractive \cite{hagen1994} and repulsive \cite{azhar2000} colloids, and the stability of binary ionic colloidal crystals \cite{leunissen2005}. Thus, characterization of $U(r)$ for colloidal particles is an important step in the process of predictably assembling target phases.

Common techniques for measuring the interaction potential of colloidal particles include colloid probe atomic force microscopy (AFM) \cite{butt2005}, surface force apparatus \cite{israelachvili1992}, total internal reflection microscopy (TIRM) \cite{bevan1999}, and optical tweezers \cite{furst2003}.  Each of these techniques allows for the direct measurement of the pair interaction potential between isolated particles and surfaces.  While these techniques provide valuable information about colloidal interactions their applications to the problems of colloidal assembly are limited because, with the exception of optical tweezers \cite{crocker1994, dufresne2007},  they do not characterize particle-particle interactions directly, which is often more relevant to the study of self-assembly of bulk colloids. Additionally, TIRM and optical tweezers are not generally applicable to refractive index matched colloidal systems since these technique require refractive index contrast for optimal performance. Often, colloidal systems that are useful for self-assembly are approximately refractive index matched, which minimizes strong attractive interactions due to van der Waals forces that tend to trigger gelation and irreversible aggregation \cite{pusey1986}.  

For colloids that are refractive-index matched, fluorescent and approximately 1$\mu$m in size, fluorescence or confocal microscopy methods can be used to characterize structure and order in such suspensions \cite{habdas2002}.  The ability to directly quantify the structure enables the use of statistical physics methods to characterize the interaction potential.  This technique involves calculating the radial distribution function, $g(r)$, and extracting the potential of mean force, $W(r)$, using the following relationship \cite{vondermassen1994},

\begin{equation}
W(r)/k_bT = -\ln[g(r)]
\label{eqnMeanForce}
\end{equation}

\noindent where $k_b$ is the Boltzmann constant and $T$ is the temperature.  $W(r)$ is a volume fraction (i.e. density) dependent measure of interaction; to determine the pair potential $U(r)$, and not simply a potential of mean force, an extrapolation to the limit of infinite dilution is required, as we will discuss in this work. These methods are complementary to the previously discussed ones because they can often be performed on the exact system that will be used in self-assembly (i.e. a bulk solution of colloids).  This basic treatment appears in various implementations in the literature.  Several groups have examined bulk solutions of colloids, capturing particle positions using video microscopy \cite{vondermassen1994} and confocal laser scanning microscopy \cite{royall2005, royall2007}.   In general, these groups calculate $g(r)$ from the microscopy data of a suspension at a particular ``dilute'' concentration and apply equation \ref{eqnMeanForce} to arrive at an ``effective'' pair potential at a finite volume fraction \cite{vondermassen1994, royall2005, royall2007}. That is, the method of these papers assumes that at dilute concentrations the potential of mean force approximates the pair potential.  Hsu et al. \cite{hsu2005} approached the problem by using bright-field microscopy.  Using a quasi-2D methodology, they captured many statistically independent images of particles interacting with one another, and then computed the 2D radial distribution function to extract the pair potential.  To corroborate their results, they employed Monte Carlo (MC) simulations to calculate $g(r$).  Wu and Bevan \cite{wu2005} took advantage of TIRM and video microscopy to capture the interaction forces between particles. In this case, colloidal pair interactions were assessed by separating out the additional contribution of the surface that is present in TIRM studies.  Finally, Lu et al. \cite{lu2008}, to support studies of gelation, parameterized short-range attractive potentials of micron-sized colloids by comparing measurements of the second virial coefficient and the cluster mass distributions to the results of MC and molecular dynamics simulations.   Overall, this class of techniques has been shown to be successful in many applications, providing results that are both consistent with theoretical behavior and simulation results.  A drawback of this basic methodology is that there is no clear definition of what is ``dilute'' and, as we discuss in detail in this paper, if we do not sample in the correct regime this method can produce results that appear qualitatively correct but are quantitatively wrong. Moreover, this paper shows that the quantitative effect of the extrapolation methodology we propose should not be underestimated and that measurements of collective structure converted to effective potentials at a small but still finite volume fraction, i.e. assuming $W(r)$  = $U(r)$ at low volume fraction, are prone to significant systematic error.

In this work, we outline a general procedure for the determination of the pair interaction potential, $U(r)$.  In this procedure we use linear regression to extrapolate the potentials of mean force, $W(r)$, at finite concentrations to infinite dilution.  We use both simulation and experiment to assess the validity of this method and provide guidelines for its use.  In section \ref{methodStat} we detail our method for determining the pair potential, providing a rational basis from statistical mechanics.  In section \ref{methodSim} we introduce our simulation model and method, and in section \ref{methodExp} we introduce our experimental method.  In section \ref{resultsSimDilute} we use simulation to determine the dilute limit, i.e. the regime where the two body forces are dominant and $W(r)$ scales linearly with volume fraction.  In section \ref{resultsSimAccuracy} we use simulation to test the accuracy of the extrapolation method. In section \ref{resultsSimPoly} we explore the role of polydispersity as it affects the potential derivation.  In section \ref{resultsExp} we apply the guidelines from section II to an experimental system of fluorescently labeled poly-methyl methacrylate (PMMA) particles stabilized by poly-12-hydroxystearic acid (PHSA) and compare to the theoretical screened Coulombic potential.   In Section \ref{conclusions} we provide concluding remarks.

\section{Methodologies}

\subsection{Method for determining $U(r)$ from the potentials of mean force \label{methodStat}}

Following Chandler \cite{chandler1987}, we can develop a rational basis for our method from statistical mechanics.  Starting with:
\begin{equation}
g(r) = e^{-\beta W(r)}
\label{remarkabletheorem}
\end{equation}
\noindent
we have a relationship between the radial distribution function, $g(r)$, and the potential of mean force, $W(r)$.  $W(r)$ represents the reversible work for the process of moving two particles from infinite separation to a finite separation of $r$.  $W(r)$ can be separated into two parts:

\begin{equation}
W(r) = U(r) +\Delta W(r)
\label{reversiblework}
\end{equation}

\noindent
where $U(r)$, the pair potential, is the reversible work to move two isolated particles to a separation of $r$ at infinite dilution, and $\Delta W(r)$ is the contribution to $W(r)$ due to the density of the system (i.e. interactions with surrounding particles in the system, not many body forces. This can additionally be thought of as moving through a potential landscape resulting from the presence of other particles).  Combining equations \ref{remarkabletheorem} and \ref{reversiblework} and taking the natural log, we arrive at:
 
 \begin{equation}
 \frac{-\ln g(r) }{\beta} = U(r) +\Delta W(r)
 \end{equation}

\noindent 
In the limit of infinite dilution (i.e. when the volume fraction, $\phi$, goes to zero) we have $\lim_{\phi \to 0} \Delta W(r) = 0$, and thus:
\begin{equation}
\lim_{\phi \to 0}  \frac{-\ln g(r) }{\beta} = \lim_{\phi \to 0} W(r) = U(r)
\label{inversionrelation}
\end{equation}

Equation \ref{inversionrelation} is the basis for our method for determining $U(r)$.  To avoid confusion, we will now refer to the potential of mean force as $W(r, \phi)$ and the radial distribution function as $g(r,\phi)$, to highlight the volume fraction dependence.  The general procedure to calculate $U(r)$ is as follows: 

\begin{enumerate}
 \item Calculate $g(r, \phi)$ at a series of finite, dilute values of $\phi$
 \item Calculate $W(r, \phi)$ from $g(r, \phi)$, using equation \ref{eqnMeanForce}, for each value of $\phi$
 \item For each value of $r$, perform a linear regression of $W(r, \phi)$ vs. $\phi$
 \item  For each value of $r$, evaluate the linear regression at $\phi = 0$ to construct an estimate of $U(r)$
 \end{enumerate}
 
\noindent
We should note that unlike other methods in the literature, this method does not rely on a single $g(r, \phi)$ but rather on the behavior of a collection of $g(r, \phi)$ data over a range of $\phi$.  In section II we will explore the benefits of using a collection of $g(r, \phi)$ data and will provide guidelines for the use of this method.  It is important to note that this methodology only applies to systems where many body forces are negligible.  It has been shown that when the electrostatic screening length is less than the interparticle separation, the assumption of pairwise additivity is well suited to predict the forces between colloids \cite{merrill}; the experimental systems we study in t his paper fit this criteria.  It is additionally important to note that the charge of the colloids may be a function of $\phi$, thus making the application of $U(r)$ to higher density systems problematic \cite{royall2006}; however, this is an issue with all methods that determine the interaction at low $\phi$ and will not be specifically addressed in this article.

\subsection{Simulation Method and Model \label{methodSim}}
To study the various aspects of colloidal pair interactions, we performed NVT Monte Carlo (MC) simulations that employ the Metropolis sampling algorithm \cite{frenkelbook}.  Simulation is powerful in this application since we explicitly know the ``true'' interaction potential (i.e. it is programmed into the simulation code), have exact control over polydispersity and volume fraction, and have no artifacts associated with identification of the particle centroid.  Thus we can assess the accuracy of the potential derivation method under well controlled, ideal conditions. We conducted simulations of spherical particles that interact via screened Coulombic interactions, modeled using the Yukawa potential  \cite{azhar2000}, given as:

\begin{equation}
\frac{U(r)}{k_bT} = 
\epsilon\frac{\exp[-\kappa \sigma(r/\sigma-1)]}{r/\sigma} 
\label{eqnYukawa}
\end{equation}

\noindent
where $\kappa$ is the inverse Debye length, $\sigma$ is the particle diameter, and $\epsilon$ is the energy at contact (dimensionless, scaled by $k_bT$).  In our simulations, particles were treated as hardcore and not allowed to overlap (i.e. $U(r)/k_bT = \infty$ when r/$\sigma <$ 1); the potential was truncated at a distance $r$ when $U(r)/k_bT \le \epsilon/60$  ~\cite{frenkelbook}.  In the limit  of very large $\kappa$, the Yukawa potential becomes very short-ranged and essentially reduces to the hard sphere potential, only capturing excluded volume \cite{auer2001}.  We used the hard sphere potential, given by equation \ref{eqnHardSphere}, to investigate this limit.

\begin{equation}
\frac{U(r)}{k_bT} = 
\begin{cases}
\infty \;, & r \le \sigma \\
0\;, & r > \sigma
\end{cases}
\label{eqnHardSphere}
\end{equation}

To model polydispersity in particle size, particle diameters were set based on a prescribed Gaussian distribution where the average particle diameter is given as $\sigma$. Particle interactions were modeled by Equations \ref{eqnYukawa} and \ref{eqnHardSphere}, however the radial separation, $r$, was adjusted based upon the deviation from the mean diameter. Specifically for a pair of particles $i$ and $j$, with diameters $\sigma_i$ and $\sigma_j$, respectively, and a center-to-center separation of $r$, the scaled radius, $r_{scaled}$, is given by:
\begin{equation}
r_{scaled} = r- 0.5(\sigma_i-\sigma)- 0.5(\sigma_j-\sigma)
\label{eqnShift}
\end{equation}
\noindent

Note that eqn \ref{eqnShift} reduces to $r_{scaled} = r$ when $\sigma_i = \sigma_j = \sigma$. Thus, the Yukawa potential for a polydisperse set of particles is given as:

\begin{equation}
\frac{U_{ij}(r)}{k_bT} = 
\epsilon\frac{\exp[-\kappa \sigma( [r + \sigma-0.5\sigma_i- 0.5\sigma_j)]/\sigma-1]}{r/\sigma} 
\label{eqnYukawaShifted}
\end{equation}

\noindent where again, $\sigma$ is the mean particle diameter. This fixes the interaction range of the potential with respect to the surface of the particle.  This assumption is reasonable since the Debye length (1/$\kappa$) is the dominant parameter in determining interaction range and is independent of particle diameter. Additionally, for simplicity, we make the assumption that the energy at contact is not a function of particle diameter, instead focusing primarily on the role excluded volume interactions play in polydispersity. Such an assumption appears reasonable; with a fixed $\epsilon$ value, a 5$\%$ increase or decrease in particle diameter only results in a $\sim$2.5$\%$ change in the charge number, Z, and a 10$\%$ change only results in $\sim$5$\%$ change in Z, as calculated using equation \ref{eqnScreenedCoulombic}.

In all cases, we used system sizes of 1000 spherical particles in a periodic box and fixed the dimensionless temperature at a value of 1. For each system we started from a random, disordered configuration and allowed the system to run for approximately one million MC timesteps, collecting data every 1000 MC timesteps.  System size effects should be minimal considering the low volume fractions studied and relatively short-ranged interaction; the box length was substantially larger than the range of interaction for all conditions studied (e.g. for $\phi$ = 0.1 the box length, $l_{box}>17 \sigma$ and for $\phi$ = 0.001, $l_{box}>80\sigma$).  Dimensionless temperature is defined as $T* = k_bT/\epsilon$. The interaction potential programmed into the simulation code will be referred to as the ``known'' interaction potential.

\subsection{Experimental Method \label{methodExp}}

\subsubsection{Synthesis and Characterization of PHSA-stabilized PMMA Particles \label{methodExpSynth}}

Non-aqueous solvents are often selected for colloidal assembly because they offer the possibility of matching the solvent and particle refractive index, thereby minimizing attractive van der Waals interactions that can interfere with assembly.  A common choice for assembly includes mixtures of cyclohexyl bromide and decalin \cite{royall2003}.  An alternative choice, particularly useful for confocal microscopy, rheology and field-assisted assembly is dioctyl phthalate \cite{solomon2006, shereda2008}.  Because of its high viscosity, this solvent is particularly compatible with the scan rate of confocal microscopy, and we use it to test our method for that reason.

Fluorescently-labeled poly(methyl methacrylate) particles stabilized by poly-12-hydroxysteric acid (PHSA) were synthesized using an adaptation of the methods of Antl et al. \cite{antl1986}, Campbell and Bartlett \cite{campbell2002}, and Pathmamanoharan et al. \cite{pathmamanoharan1997}, as discussed by Solomon and Solomon \cite{solomon2006}.  Particles were labeled using Nile red dye.  Previous work has shown that PHSA-PMMA colloids in cyclohexyl bromide (CHB) and decalin are charged \cite{royall2003}.  Shereda \textit{et al.} also reported particle charging for these particles in DOP\cite{shereda2008}.  Such colloidal systems have been modeled by the Yukawa pair potential (equation \ref{eqnYukawa}) where we explicitly define:

\begin{equation}
\epsilon = \frac{Z^2\lambda_B}{\sigma_p k_B T (1+\kappa \sigma_p/2)^2}
\label{eqnScreenedCoulombic}
\end{equation}

\noindent where $Z$ is the charge number, $\lambda_B$ is the Bjerrum length (11 nm for the system studied here \cite{russel1989}), and $\kappa$ is the inverse Debye length \cite{leunissen2005}. 

Parameters in equations \ref{eqnYukawa} and \ref{eqnScreenedCoulombic}, estimated by experiment, are reported in Table \ref{particlechar}.  $\kappa^{-1}$ was estimated from conductivity measurements.  Briefly, solutions of dioctyl phthalate (DOP; Sigma-Aldrich, used as received) containing 10 $\mu$M and 2mM of tetrabutylammonium chloride (TBAC) salt were prepared and their conductivity measured with devices (Model 1154 Precision Conductivity Meter, Emcee Electronics, Venice, Florida or Model EW-01481-61, Cole-Parmer, USA) whose performance was verified by measurements with known standards.  As per the method of reference \cite{royall2003}, using WaldenÕs rule \cite{fuoss1959,walden1906,walden1906b} (viscosity of DOP =  71 mPa s at 25¡C \cite{dean1999} and reference ion mobilities in water at 25C (as taken from reference \cite{royall2003} TBA$^+$ ion mobility is 19.4 cm$^2$ S mol$^{-1}$ and Cl$^-$ is 76.3 cm$^2$ S mol$^{-1}$), the ion concentration was determined and calculated using equation \ref{eqnDebye}:

\begin{equation}
\kappa=\sqrt{8\pi\lambda_B\rho_i}
\label{eqnDebye}
\end{equation}

\noindent
{where $\rho_i$ is the density of the cation or anion \cite{royall2003}. The Debye length of salt-free DOP was not determined in this way because of the uncertain identification of the mobile ions in this solvent. Thus, in this case, a one-parameter fit to the extrapolated pair potential (as discussed in the results) was made and a value of 440 nm was obtained. This value is included in the table for completeness.   This Debye length is qualitatively consistent with literature reports \cite{bautista2007} of the very low conductivity ($\sim$ 2.3 x 10$^{-9}$ S/m) of dioctyl phthalate.

%The conductivity of salt-free DOP was also below the resolution of our instrument.  In this case, a one-parameter fit to the extrapolated pair potential (as discussed in the results) was made and a value of 444 nm was obtained.  This value is included in the table for completeness.  This Debye length corresponds to an effective conductivity of $2.85x10^{-4}$ $\mu$S/cm.  Although this value is higher than published reports of the conductivity }, it was consistent with a measurement performed by calculating the resistance of DOP in a Patch-1U Model cell ($5.35x10^{-4}$ $\mu$S/cm).

The charge number of the PHSA-stabilized PMMA particles was determined from measurements of their electrophoretic mobility (Zetasizer Nano Series, Malvern, United Kingdom).  Particles were prepared as 1 vol percent solutions in pure DOP solvent (i.e. [TBAC] = 0), [TBAC] = 10$\mu$M, and [TBAC] = 2mM.  Solutions were placed in a dip cell designed for non-aqueous solvents.  A voltage of 50mV was applied to each sample and three independent samples were studied.  Using the calculated Debye length and measured mobility, the zeta potential was determined by the method of O'Brien and White \cite{obrien1978}.  The charge number on the particles was determined from: 

\begin{equation}
Z=\frac{q}{e}=\frac{4\pi \epsilon_0\epsilon_r(\sigma/2)\zeta(1+\kappa\sigma/2)}{e}
\end{equation}

\noindent where $q$ is the particle charge, $e$ is the charge on an electron, $\epsilon_0$ is the permittivity in a vacuum, $\epsilon_r$ is the relative dielectric constant of DOP (~5.10 \cite{crchandbook}), and $\zeta$ is the zeta potential \cite{russel1989}. 

The distribution of particle diameters, determined from scanning electron microscopy (SEM) analysis of 250 particles, is plotted in Figure \ref{particlesizedist}.  From the best fit of the normal distribution, we determined the mean diameter to be 951nm $\pm$ 38nm.  Because PMMA colloids may swell in organic solvents \cite{kogan2008}, we compared the SEM diameter to direct measurements of the colloid size in solution.  To perform the comparison, we prepared a sample at a very high TBAC salt concentration of 50mM to induce aggregation of the particles in the solvent.  A CLSM image volume was acquired and the separation between particle pairs was computed by image processing per the method described subsequently in section \ref{methodImagesProcessing}.  We found the particles swell approximately 5$\%$ to 1001 $\pm$ 30 nm.  This small change in diameter does not have a major effect on our measurements or $U(r)$.  For consistency, we will report our findings based on the particle diameter measured by the CLSM method, since our experiments were done in DOP solvent.

\begin{figure}[ht]
\includegraphics[width=3.25in]{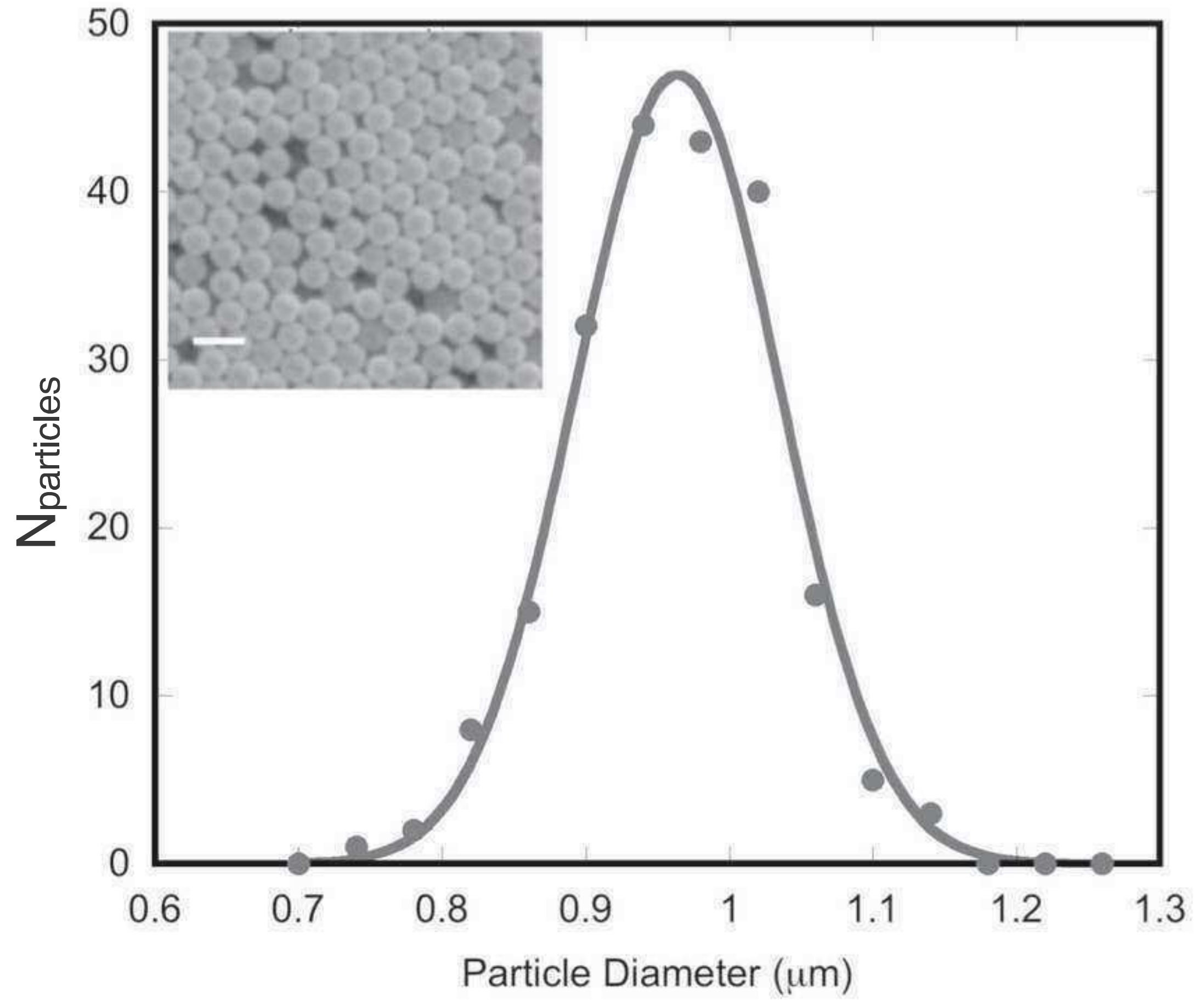} 
\caption{Distribution of PHSA-PMMA particles diameters with fitted Gaussian curve shown as solid line; particles have a diameter of 951 $\pm$ 38nm.  Inset is an SEM image of PHSA-stabilized PMMA particles used to generate the distribution; scale bar represents 2$\mu$m.}
\label{particlesizedist}
\end{figure}

\begin{table}[htdp]
\begin{center}
\begin{tabular}{c c c c c c}
DOP containing & $\kappa^{-1}$ (nm) & $\zeta$ (mv) & $Z_{CLSM}$ & $\epsilon_{CLSM}$ & $\kappa\sigma_{CLSM}$ \\
\hline
No TBAC & 440 & -28 &-100 & 27 & 2.2 \\
10 $\mu$M TBAC & 250 & -30 & -160 & 31 & 4.0 \\
2 mM TBAC & 49 & -32 & -630 & 35 & 20 \\
\hline
\label{particlechar}
\end{tabular}
\caption{Electrokinetic measurements for the three experimental systems studied.}
\end{center}
\label{default}
\end{table}%

\subsubsection{Sample preparation and image volume collection}
In this work, we explored three different systems where the solvent and particles are approximately refractive index (RI) matched.  The PMMA-PHSA particles (RI = 1.489) were dispersed in DOP (RI = 1.485) containing no salt, 10$\mu$M TBAC salt or 2mM TBAC salt.  Samples were prepared at nominal volume fractions of $\phi$ = [0.005, 0.01, 0.015, 0.02, 0.03, 0.04, 0.05]; however, precise estimation of the volume fraction was made from the results of quantitative image processing discussed in section \ref{methodImagesProcessing}.  The samples were initially mixed and then allowed to equilibrate for 24 hours.  Samples were then gently remixed and subsequently transferred to glass specimen vials (outer diameter, O.D. = 12mm) that were adhered to a microscope cover slip using ultraviolet bonding glue (Dymax Corporation, United States).  The cover slip was attached to a 35mm O.D. glass ring made from Pyrex standard wall tubing.   To assess the stability of this colloidal system, samples prepared at $\phi$ = 0.005 in the DOP solvent were monitored for 24 hours for signs of phase instability.  No aggregation or phase instability was observed. 

Sample imaging was performed on a Leica TCS SP2 confocal microscope (Leica Microsystems, Wetzlar, Germany).  A 100x oil immersion objective with numerical aperture 1.4 was used.  The particles were dyed with Nile red (Sigma-Aldrich, United States) and were excited with a green neon (GreNe) laser beam ($\lambda_0$ = 543nm).  Emission from 550nm to 650nm was collected.  To avoid possible effects of sample boundaries on particle configurations, all points in the image volumes were located at least 20$\mu$m from any boundary of the specimen vial.  Stacks of 247 images with a resolution of 512 x 512 pixels were obtained and processed to extract particle centroids, as described in the next section.  Images were acquired with a spatial resolution of 69.2 x 69.2 nm/pixel in the objective plane and an axial separation of 81.4 nm.   Thus, the size of the image volume was 35 x 35 x 20 $\mu m^3$. 

\subsubsection{Image Processing \label{methodImagesProcessing}}
To identify particle locations, we used image processing algorithms based on the work of Crocker and Grier \cite{crockergrier1996} expanded to 3D systems \cite{varadan2003} as discussed in Dibble et al. \cite{dibble2006}.   First, a Gaussian filter was applied to the 3D image volume. Second, particle centers were identified using a local brightness maximum criterion.  That is, a voxel was identified as a candidate centroid if it was the brightest within a local cubic region of half-width $w$.  For our systems and imaging conditions, $w$ = 7, which corresponded to approximately 485 nm.  Finally, using the moments of the local intensity distribution, particle positions were refined to subpixel accuracy \cite{crocker1996}.  For our setup, the accuracy was calculated to be $\pm$35 nm in the $x-y$ plane and $\pm$45 nm in the $z$ direction in reference \cite{dibble2006}.  $g(r,\phi)$ was calculated using the centroid locations.  We assessed the accuracy of the image processing algorithm by examining composite images for which centroid locations were overlaid on the fluorescence images; the algorithm was found to identify nearly all of the particle centroids.  We also validated the centroid calculation for particles at close contact, finding no evidence of erroneous behavior, as has been observed for other systems \cite{ramirezsaito2006, baumgartl2006}. For our fluorescently labeled system, CLSM provided a strong contrast between particle and solvent, and does not produce an Airy pattern; thus we meet both conditions shown to be essential for accurate centroid determination outlined in reference \cite{ramirezsaito2006}.

\section{Simulation Results}
\subsection{Determining the Dilute Regime \label{resultsSimDilute}}

The general method for calculating the pair potential involves linearly regressing $W(r,\phi)$ at finite values of $\phi$, and evaluating the regression at $\phi$ = 0.  In order for this to be valid, we must be in a range where $W(r,\phi)$ scales linearly with $\phi$ (i.e. the regime where only 2-body forces are dominant and there are no strong volume fraction effects).  Unfortunately, there is no general rule for what value of $\phi$ constitutes this dilute regime; the crossover between ``dilute'' and ``concentrated'' will be a function of the strength and relative range of the interaction.  To determine the crossover, $\phi_{max}$, we performed a series of simulations using the hard-sphere and Yukawa potentials over a range of $\phi$.  Specifically for the Yukawa potential, we performed simulations with $\epsilon$ = 3 and $\kappa \sigma$ = (5.0, 3.0, 1.5, 1.0, 0.5, 0.25), $\epsilon$ = 50 and $\kappa \sigma$ = (5.0, 3.0, 1.5, 1.0), and $\epsilon$= 100 with $\kappa \sigma$ = (5.0 ,3.0, 1.5, 1.0) at a variety of values of $\phi$.  This allows us to explore the dependence of $\phi_{max}$ on both interaction range and interaction strength.  

\begin{figure}[ht]
\includegraphics[width=2.75in]{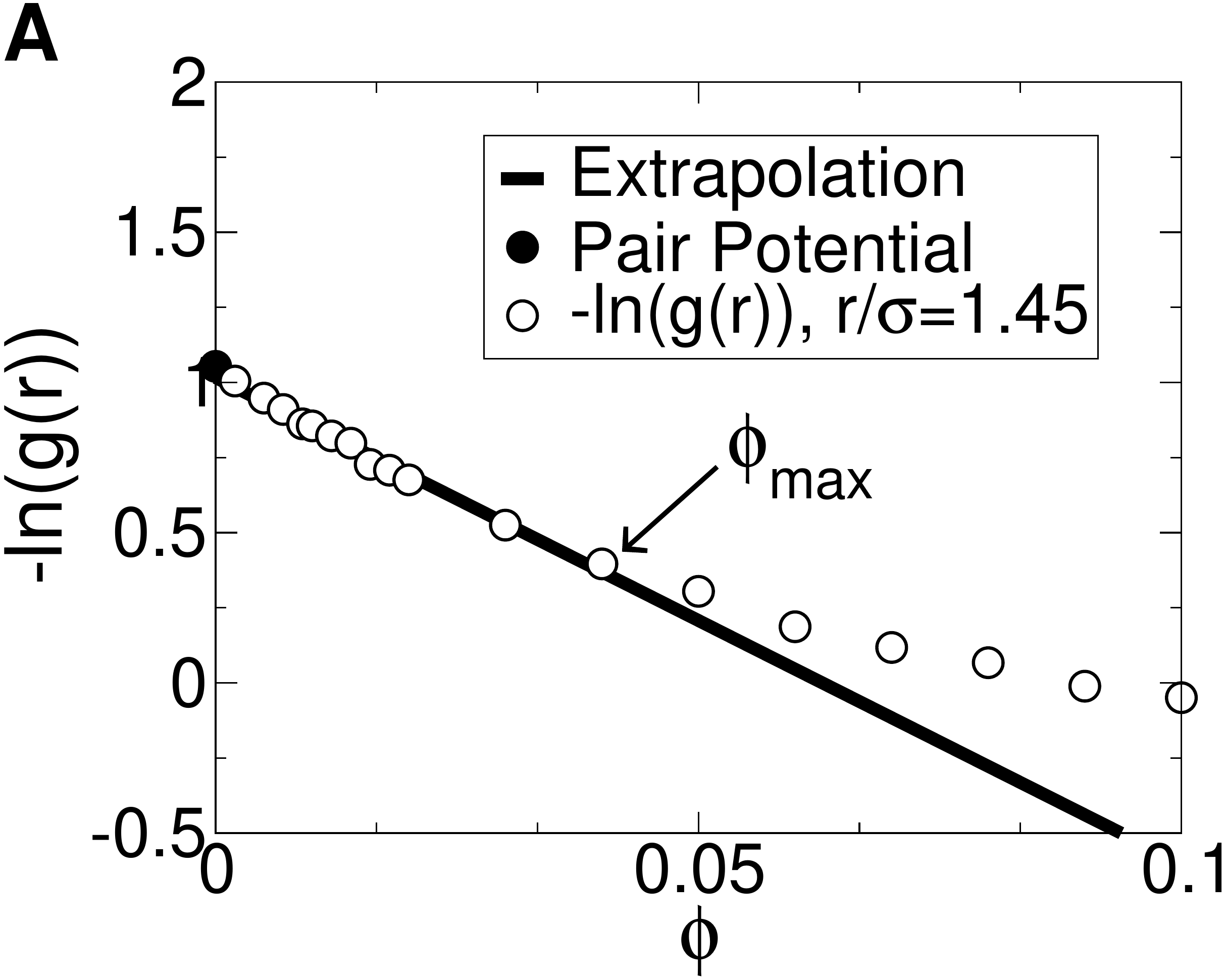} 
\includegraphics[width=2.75in]{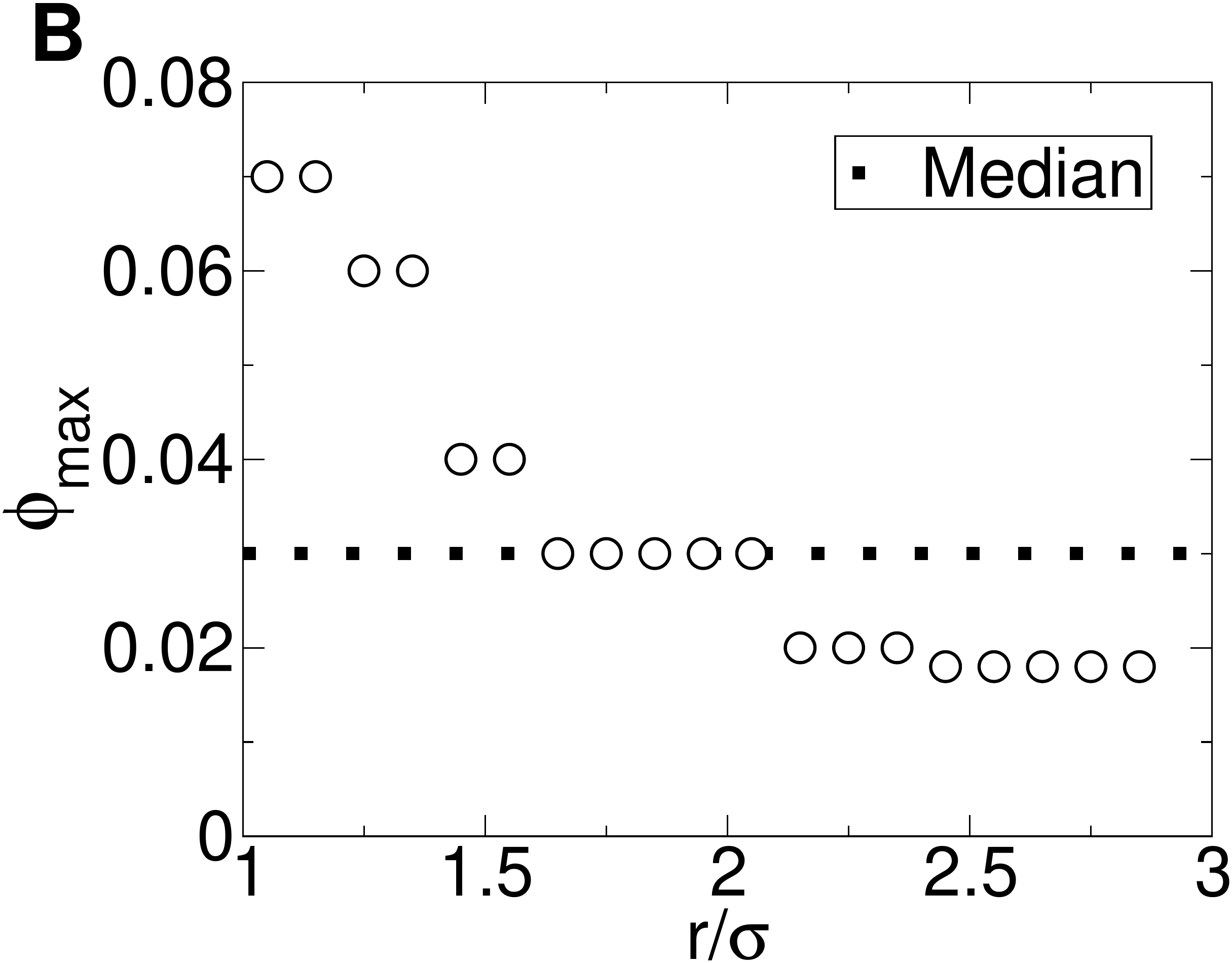} 
\caption{For a Yukawa system with $\epsilon$ = 3, $\kappa \sigma$ = 1.5, (A) determination of the dilute regime where  is the crossover, and (B)  vs.  where the median value appears as the dotted line.}
\label{figurePhiMaxvR}
\end{figure}

To determine $\phi_{max}$, the maximum volume fraction for which $W(r, \phi)$ scales linearly with $\phi$, we performed the following procedure.  For each value of $r/\sigma$, we plotted  $-\ln[g(r,\phi)]$ (i.e. $W(r,\phi)$) as a function of $\phi$; we included the pair potential as the value that occurs at $\phi$  = 0.  We then determined, by eye, the approximate crossover between linear and non-linear behavior.  We then performed a linear regression on the subset of the data that appeared linear and fine-tuned our results.  We looked at both the residual of the fitting of the data and compared between the known  (i.e. the potential programmed into the simulation) and regressed value at $\phi$ = 0 to determine the appropriate value of $\phi_{max}$.   Figure \ref{figurePhiMaxvR}A plots an example of the crossover between ``dilute'' and ``concentrated'' regimes showing a region with distinct linear behavior; the extrapolation is also plotted showing near perfectly agreement with the known interaction potential value. This procedure was applied to all values of $r/\sigma$, creating a data set for each potential that is of the form of $\phi_{max}$ vs. $r/\sigma$; an example data set is plotted in Figure \ref{figurePhiMaxvR}B.  We should note that in Figure \ref{figurePhiMaxvR}B it is clear that $\phi_{max}$ decreases as we increase $r/\sigma$ and it is not simply a single value for a given pair interaction. The dependence of $\phi_{max}$ on $r/\sigma$ for a fixed $\epsilon$ and $\kappa\sigma$ can be understood by thinking in terms of the ``effective'' diameter of the particle. That is, if we were to increase the diameter of our particle, we would need to increase our system volume to maintain the same $\phi$.  The quantity $r/\sigma$ can be considered to be proportional to the ``effective'' diameter of the particles, thus, as we increase $r/\sigma$ we need to also increase our system volume to maintain the same effective $\phi$, as we see in Figure \ref{figurePhiMaxvR}B (i.e. increasing system volume is analogous to a decrease $\phi$ since  $\phi$ is calculated using the excluded volume diameter, $\sigma$, not an effective diameter).  As such, we calculated the median value of $\phi_{max}$, as this should be representative of the data and not biased by the large values of $\phi_{max}$ at small $r/\sigma$.

\begin{figure}[ht]
\includegraphics[width=3.5in]{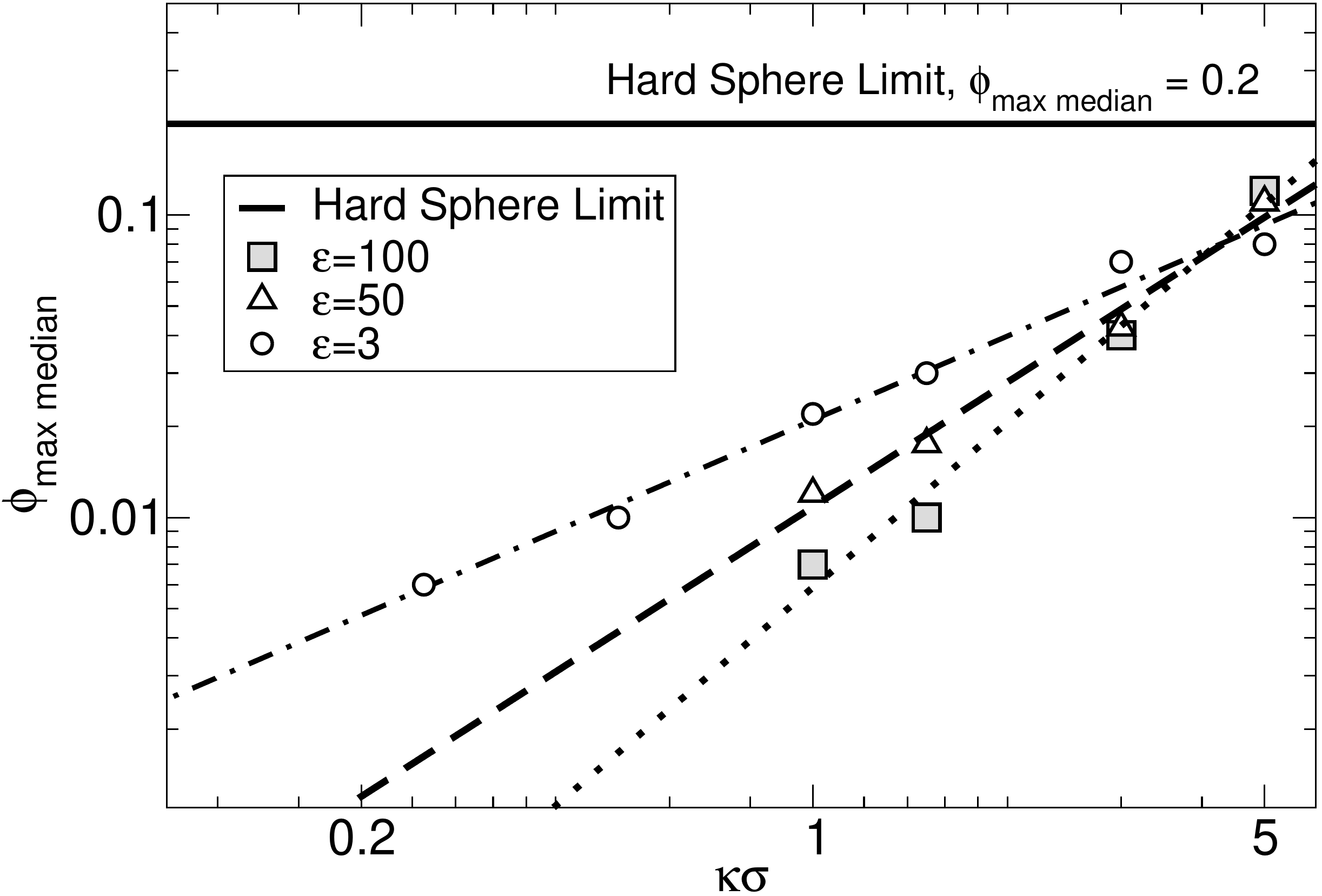} 
\caption{The median value of $\phi_{max}$ vs. $\kappa \sigma$ showing the approximate crossover between concentrated and dilute regimes. Data for each $\epsilon$ value are fitted with a power law best fit to guide the eye.}
\label{figurePhiMaxScaling}
\end{figure}

The median values are summarized in Figure \ref{figurePhiMaxScaling}.  The data sets were grouped by $\epsilon$ and fitted with a power law to guide the eye.  We see that as $\kappa \sigma$ decreases  (i.e. our potential becomes longer ranged with respect to particle diameter), $\phi_{max,median}$ decreases.  Additionally, as $\epsilon$ increases, we find that $\phi_{max,median}$ also decreases, however not as rapidly as the trends with $\kappa \sigma$.  We also see that for the hard-sphere system, $\phi_{max,median}$ = 0.2, and this can be taken as an upper limit for all repulsive systems; $\phi_{max,median}$ for hard-spheres was calculated using the same procedure as the Yukawa systems and corresponds to a state where $\kappa$ = $\infty$, thus it is plotted as a bounding line.  It is clear that as the range of the potential increases, we must sample at increasingly lower volume fractions. 

Figure \ref{figurePhiMaxScaling} can be used as a rough guideline to determine the appropriate regime to collect data for the calculation of $U(r)$.  While this plot depends on \textit{a priori} knowledge of $\kappa \sigma$ and $\epsilon$ -- the values we ultimately wish to calculate -- we only need a rough estimate of the screening length to use this plot.  As previously stated, the median value of $\phi_{max, median}$ depends predominantly on $\kappa \sigma$ which can be roughly estimated with minimal effort by measuring at a low value $g(r,\phi)$ (e.g. $\phi$ = 0.01 is satisfactory for the systems presented in Figure \ref{figurePhiMaxScaling}). In practice, the values of $\phi_{max}$ should be assessed for each system studied using the methodology previously described; Figure \ref{figurePhiMaxScaling} should be used as a rough guide to approximate where this regime occurs, to avoid collecting unnecessary data.

\subsection{Accuracy of the Extrapolated Potential \label{resultsSimAccuracy}}

We applied the extrapolation method to our Monte Carlo simulations and found excellent agreement between the extrapolated potentials and the pair potentials programmed into the simulation; note in these cases we treated the system as monodisperse with respect to particle diameter. In Figure \ref{figureExtrapolated} we plotted the known potential and the extrapolated potential for an example system with $\epsilon$=3 and $\kappa \sigma$ =1.5. We extrapolated $W(r,\phi)$ for 11 values of $\phi$ ranging from  0.002 to 0.03, using the median value of $\phi_{max,median}$ = 0.03 as the maximum cutoff for the regression; we did not take into account the $r/\sigma$ dependence of $\phi_{max}$ for this test.  We see that the extrapolated potential is virtually identical to the known potential, as shown in Figure \ref{figureExtrapolated}.  To parameterize the accuracy of the extrapolated potential, we calculated the magnitude of the difference between the test and known potentials, normalized by the magnitude of the two potentials (represented as vectors), 

\begin{equation}
M = \frac{\sum_i^{max}|U_{test}(r_i) - U_{known}(r_i)|}{\sum_i^{test}|U_{test}(r_i)|+\sum_i^{max}|U_{known}(r_i)|}
\end{equation}

\noindent
where $max$ is the total number of discrete points considered; in this measure an ideal match has $M$ =  0 and the maximum difference has $M$= 1.  We used $M$ rather than the relative error since $M$ is always normalized between 0 and 1 for all potentials and number of datapoints considered.  Additionally, by its construction, $M$ is weighted with respect  to the total magnitude of both vectors we compare, thus we avoid any erroneously large errors associated with the difference between relatively small values (e.g. $U(r)$ = 0.05 and $U(r)$ = 0.01 are nearly indiscernible on the energy scale we consider here, however their relative error is 4, whereas the difference between $U(r)$ = 10 and $U(r)$ =  6 would be clearly evident, yet the relative error is much less with a value of 0.67). The difference between the extrapolated potential and known pair potential in Figure \ref{figureExtrapolated} was calculated to be $M$=0.023.

The advantage of using the regression method over a single ``dilute'' concentration is also highlighted in Figure \ref{figureExtrapolated}.  We plotted $W(r,\phi)$ for two different volume fractions, $\phi$ =0.05 and 0.005.  For a value of $\phi$ = 0.05, a relatively low-density system, we capture the gross behavior, however the actual values deviate substantially from the known potential.  $W(r, 0.05)$ significantly undershoots the known potential and any values of $\epsilon$ and $\kappa$ extracted from this plot would be misleading; for $W(r, 0.05)$, $M$ =0.497.  We also plotted $W(r,\phi)$ for $\phi$ = 0.005, a very low density sample, and find this to be a reasonable approximation to the known potential with $M$ = 0.063. 
\bigskip
\bigskip

\begin{figure}[ht]
\includegraphics[width=3.5in]{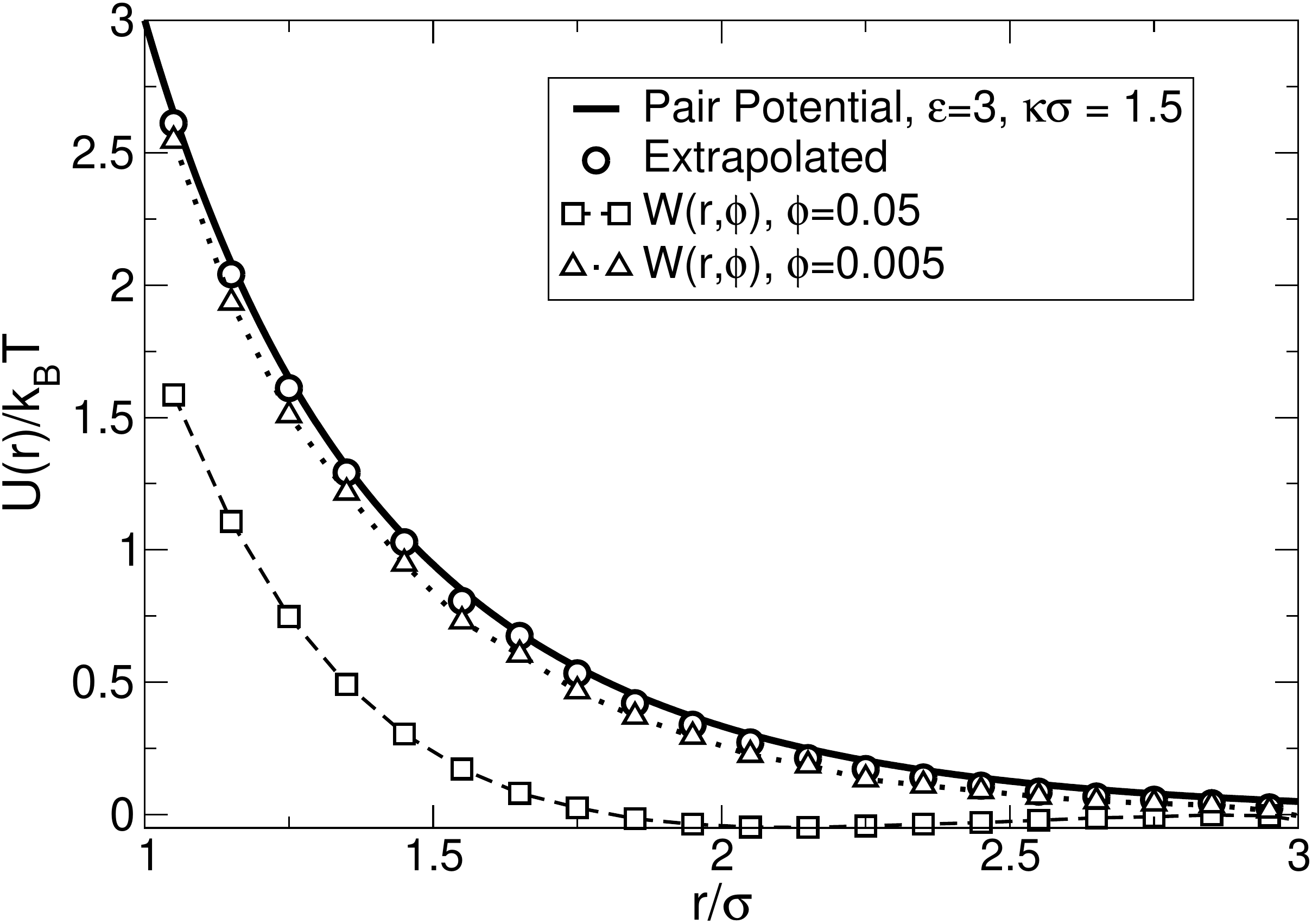} 
\caption{For a Yukawa system with $\epsilon$ = 3 and  $\kappa \sigma$ = 1.5, the pair potential and extrapolated potentials are plotted showing excellent agreement.  The extrapolated potential was calculated from 11 values of $\phi$ ranging from 0.002-0.03. $W(r,\phi)$ is also plotted for $\phi$ = 0.05 and 0.005. $M$ = 0.023 for the extrapolated potential, $M$ = 0.497 for $W(r,0.05)$, and $M$ = 0.063 for $W(r,0.005)$}
\label{figureExtrapolated}
\end{figure}

Without prior knowledge of the known potential -- besides the notion it should be repulsive -- it would difficult to assess if $W(r, 0.05)$ or $W(r,0.005)$ are good approximations of the pair potential.  In order to assess whether the potential of mean force is reasonable, we would need to follow a procedure similar to that outlined in reference \cite{vondermassen1994}; we would need to collect additional data at lower $\phi$ values and see if $g(r, \phi)$, or alternatively $W(r,\phi)$, changed substantially as $\phi$ was decreased \cite{vondermassen1994}.  In effect, we would need to collect, analyze, and calculate additional data that would not necessarily be directly used to calculate the approximation of the pair potential. Therein lies a substantial benefit of the extrapolation method.  In addition to calculating a more accurate approximation of the pair potential, we use nearly all data collected to calculate it, thereby increasing our overall confidence in the result; this procedure also provides a built-in ``check'' regarding the appropriateness of the derived potential in terms of its dependence on $\phi$ (i.e. if $W(r,\phi)$ does not scale linearly with $\phi$, we are not dilute enough).

In Figure \ref{figureE3Summary} we plotted the extrapolated potentials calculated for $\epsilon$ = 3, and $\kappa \sigma$ = (5.0, 3.0, 0.5, 0.25) to explore the behavior of this method as a function of screening length.  As in Figure \ref{figureExtrapolated}, we ignored the $r/\sigma$ dependence of $\phi_{max}$ and used only the median value.  The extrapolation method worked well in all cases.  The average magnitude of the difference between the known pair potential (i.e. the potential coded into the simulation) and extrapolated potential for $\epsilon$ = 3 and $\kappa \sigma$ = (5.0, 3.0, 0.5, 0.25) is $M$ = (0.028, 0.021, 0.046, 0.087), respectively.  As the range of the potential increases (i.e $\kappa \sigma$ becomes smaller), the method slightly underestimated the known value of the pair potential at larger $r/\sigma$ values; this was most evident for  $\kappa \sigma$ = 0.25.  As the range of the potential increases, the median of $\phi_{max}$, or any single value, is less representative of the data on a whole, and a more careful treatment of the regression should be applied (i.e. $\phi_{max}$ should be determined for each value of $r/\sigma$).  However, the deviations are still well within the expected experimental error.
\bigskip

\begin{figure}[ht]
\includegraphics[width=3.5in]{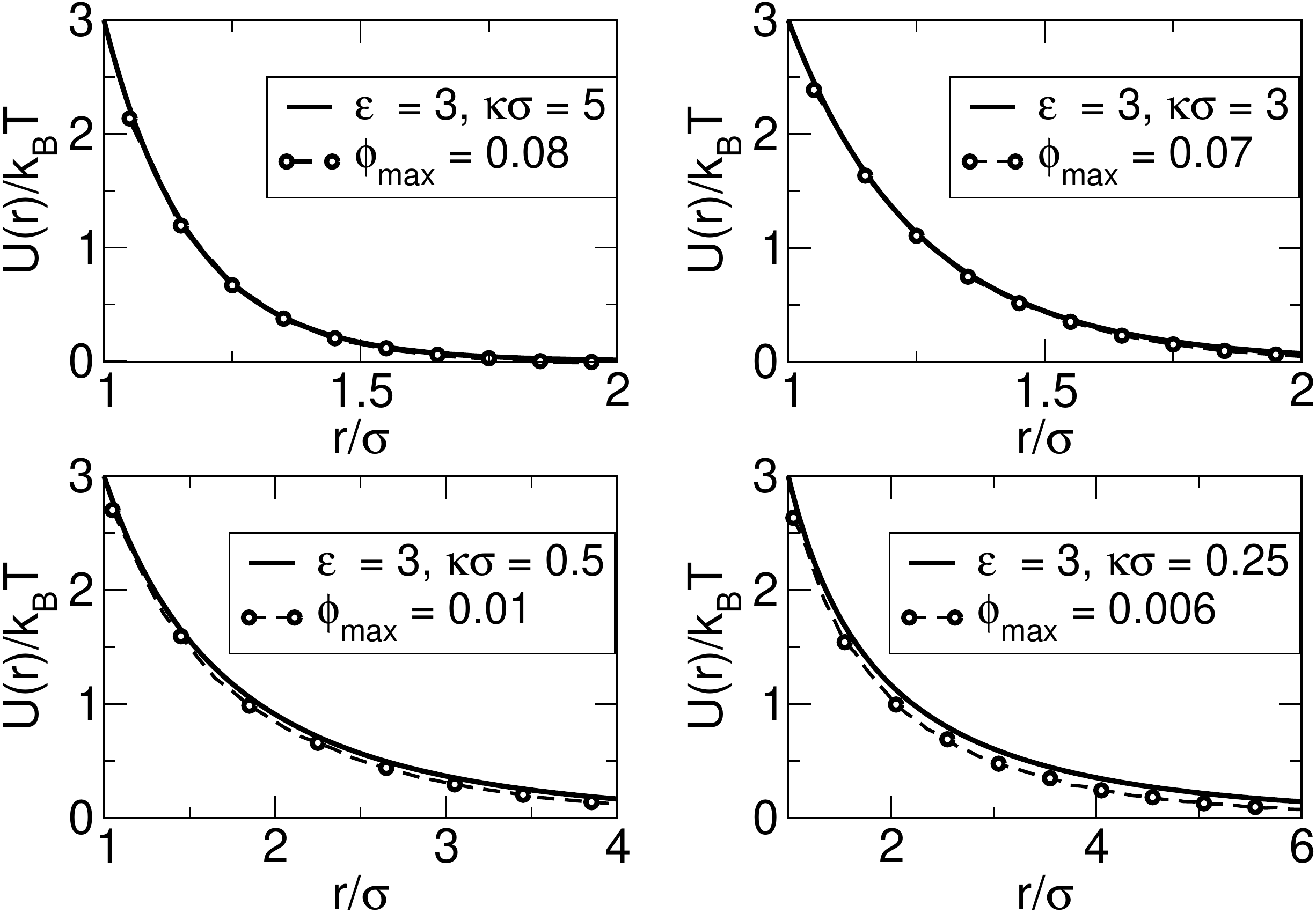} 
\caption{Comparison between the known and extrapolated pair potentials.  In all cases, the known potential is shown as a solid black line, and the extrapolated potential is shown as circles.  In all cases, the extrapolated potential matches the known potential showing only minor deviations as the range of the potential is increased (i.e. $\kappa \sigma$ decreases).}
\label{figureE3Summary}
\end{figure}

In Figure \ref{figureEpsilonSummary} we explored the impact of $\epsilon$ on the extrapolation method, plotting the results for system where $\epsilon$ = 50 and  $\kappa \sigma$ = (3.0, 1.0) and $\epsilon$ = 100 and $\kappa \sigma$ = (3.0, 1.0).   In all cases, the extrapolation method matched the known potential well.  The average magnitude of the difference between the known and extrapolated potentials for $\epsilon$ = 50 and $\kappa \sigma$  = (3.0, 1.0) is $M$ = (0.038, 0.087), respectively and for $\epsilon$ = 100 and $\kappa \sigma$ = (3.0, 1.0) $M$ = (0.022, 0.024), respectively. However, we see that the extrapolation method does not resolve values of $U(r)/k_BT$ greater than $\sim$10.  This is not a failing of the method, but rather a consequence of the nature of the strong interactions in these cases.  That is, as the strength of the potential is increased, the likelihood of two particles coming into very close contact is lowered.  %In the Metropolis algorithm used in our Monte Carlo simulations, we accept or reject particle moves based on a Boltzmann probability; particle moves that raise the potential energy (such as moving into the strongly repulsive region of the potential) have a low probability of being accepted.  Or alternatively, in a dynamical system, particle motion -- and therefore kinetic energy -- is linked to the temperature, specifically $\langle KE \rangle = \frac{3}{2}k_bT$ .  Thus, in dimensionless units, a system at T* = 1.0 will have an average per particle energy of 1.5 making it unlikely for any given particle to have an energy much greater than 10, which would be necessary to overcome the strength of $U(r)/k_BT$; thus $g(r,\phi)$ is unlikely to show any particle contacts in regimes where the potential is very strong for low $\phi$.
\bigskip
\bigskip
\bigskip

\begin{figure}[ht]
\includegraphics[width=3.5in]{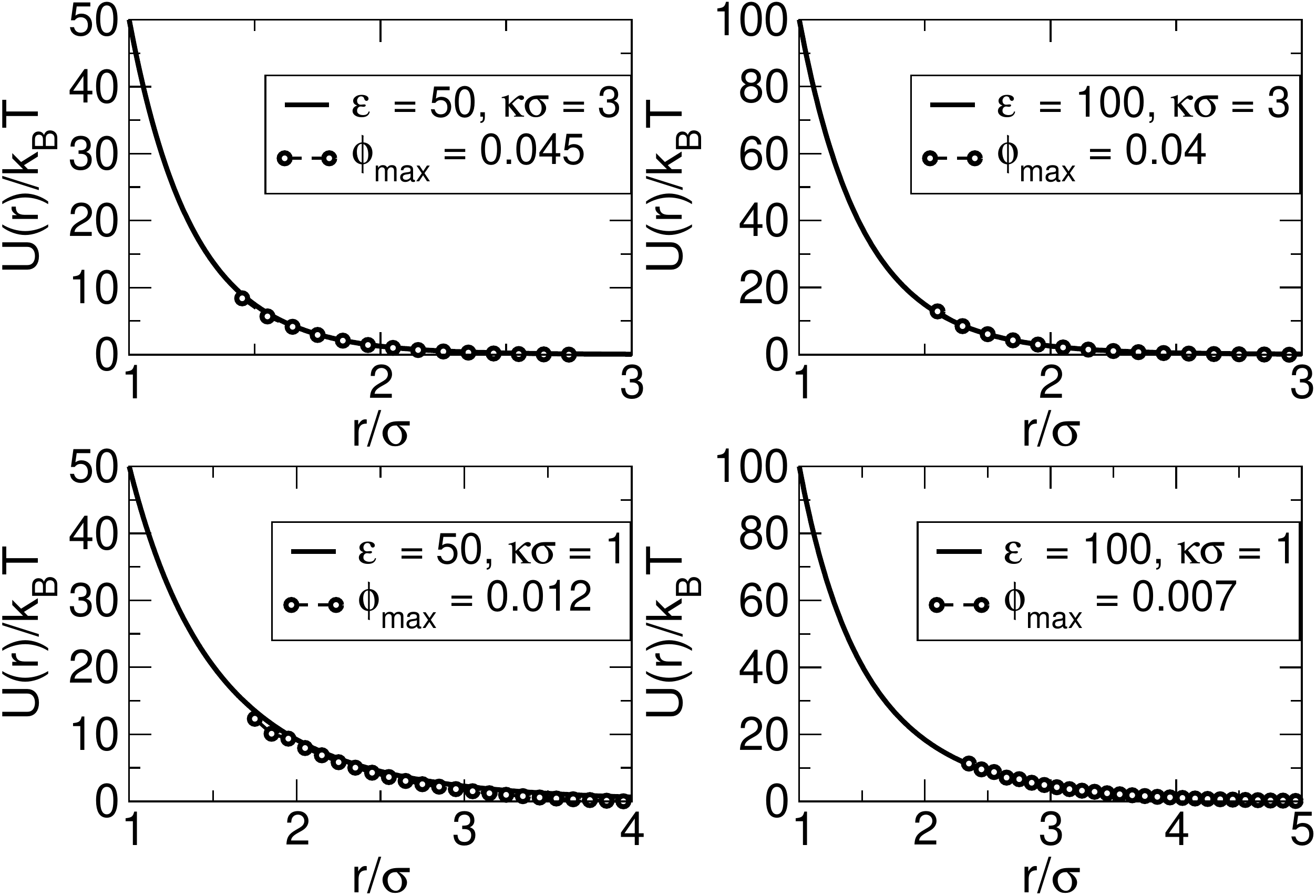} 
\caption{Comparison between the known and extrapolated pair potentials.  In all cases, the known potential is shown as a solid black line, and the extrapolated potential is shown as circles.  The extrapolated potential matches the known potential well in all cases, however, we can only resolve the potential on the order of $\sim$10 $k_BT$.}
\label{figureEpsilonSummary}
\end{figure}

\subsection{Effect of Polydispersity \label{resultsSimPoly}}
In sections \ref{resultsSimDilute} and \ref{resultsSimAccuracy} we treated our model colloidal particles as ideal and thus ignored polydispersity in particle diameter.  In practice, experimental systems will contain particles with a range of diameters, often distributed with Gaussian behavior, as we observed in Figure \ref{particlesizedist}. Polydispersity is defined as, P = (100 $\%$)(standard deviation of particle diameter)/(average particle diameter). The experimental systems we explored in this work (see section \ref{methodExp}) have polydispersity levels of P $\sim$ 4$\%$ and this level is not atypical of colloidal systems for self-assembly.  Polydispersity has been shown to have a large effect on phase behavior, for instance creating a phase separated crystal under high density when P $>$ 8$\%$ \cite{sear1998}. However, since we are in the dilute regime, the effects of polydispersity may be muted. In our simulations, we initialized our particle diameters based on a Gaussian distribution that satisfied the given level of polydispersity we wish to study; our methodology was previously described in section \ref{methodSim}.

We simulated three Yukawa systems with $\epsilon$ = 3 and $\kappa \sigma$ = 5,  $\epsilon$ = 3 and $\kappa \sigma$ = 10, and $\epsilon$ = 43 and $\kappa \sigma$ = 20, each for P = (0 $\%$, 5 $\%$, 10 $\%$); the extrapolations are shown in Figure \ref{figurePotentialPoly}.  For $\epsilon$ = 3 and $\kappa \sigma$ = 5, polydispersity has a minimal effect on the extrapolation if we consider the extrapolated $U(r)$ for $r/\sigma \ge$ 1, i.e  we do not consider $r/\sigma$ values that are less than the separation between two average diameter particles.  Specifically, for P = (0,5,10)$\%$ we find $M$ = (0.028, 0.033, 0.049) respectively when calculating $M$ for $r/\sigma \ge$ 1.  

Increasing $\kappa \sigma$ to 10 makes the potential shorter ranged and steeper.  In this case, polydispersity appears to have more of an impact; specifically, P =10$\%$ undershoots the known potential.  However, as we saw for $\kappa \sigma$= 5, truncating $U(r)$ at $r/\sigma \ge$ 1 results in nearly identical  extrapolated pair potentials at all levels of polydispersity. For P=(0,5,10)$\%$ we find $M$ = (0.030, 0.041, 0.080) respectively when calculating $M$ for $r/\sigma \ge$ 1.  

For a system with $\epsilon$ = 43 and $\kappa\sigma$ = 20, the range of the potential, with respect to the surface of the particle, is roughly equal to the case where $\epsilon$ = 3 and $\kappa\sigma$ = 10, however it increases more rapidly at small $r/\sigma$.  This rapid increase in interaction strength results in a system where the diameter of the particle effectively increases (since it is unlikely for particles to interact when $U(r)/k_bT > 10 $). Hence, the effective interaction is actually shorter-ranged than $\epsilon$ =3 and $\kappa \sigma$ = 10 (i.e. the interaction can be thought of as shorter-ranged and radially shifted outwards from the surface of the particle).  This is particularly evident as even the P =$0\%$ system has trouble resolving the known pair potential well at small $r/\sigma$.  The effect of polydispersity seems more pronounced than the other systems, undershooting the known potential in both the 5$\%$ and 10$\%$ cases.  In this case, simply truncating our extrapolation at $r/\sigma \ge$ 1 is not sufficient to minimize the impact of polydispersity.  For P=(0,5,10)$\%$ we have $M$ = (0.302, 0.47, 0.59) respectively when calculating $M$ for $r/\sigma \ge$ 1.  However, we note that for $r/\sigma \ge 1.125$, we are able to resolve the potential reasonably well for all levels of polydispersity.

\begin{figure}[ht]
\includegraphics[width=3.0in]{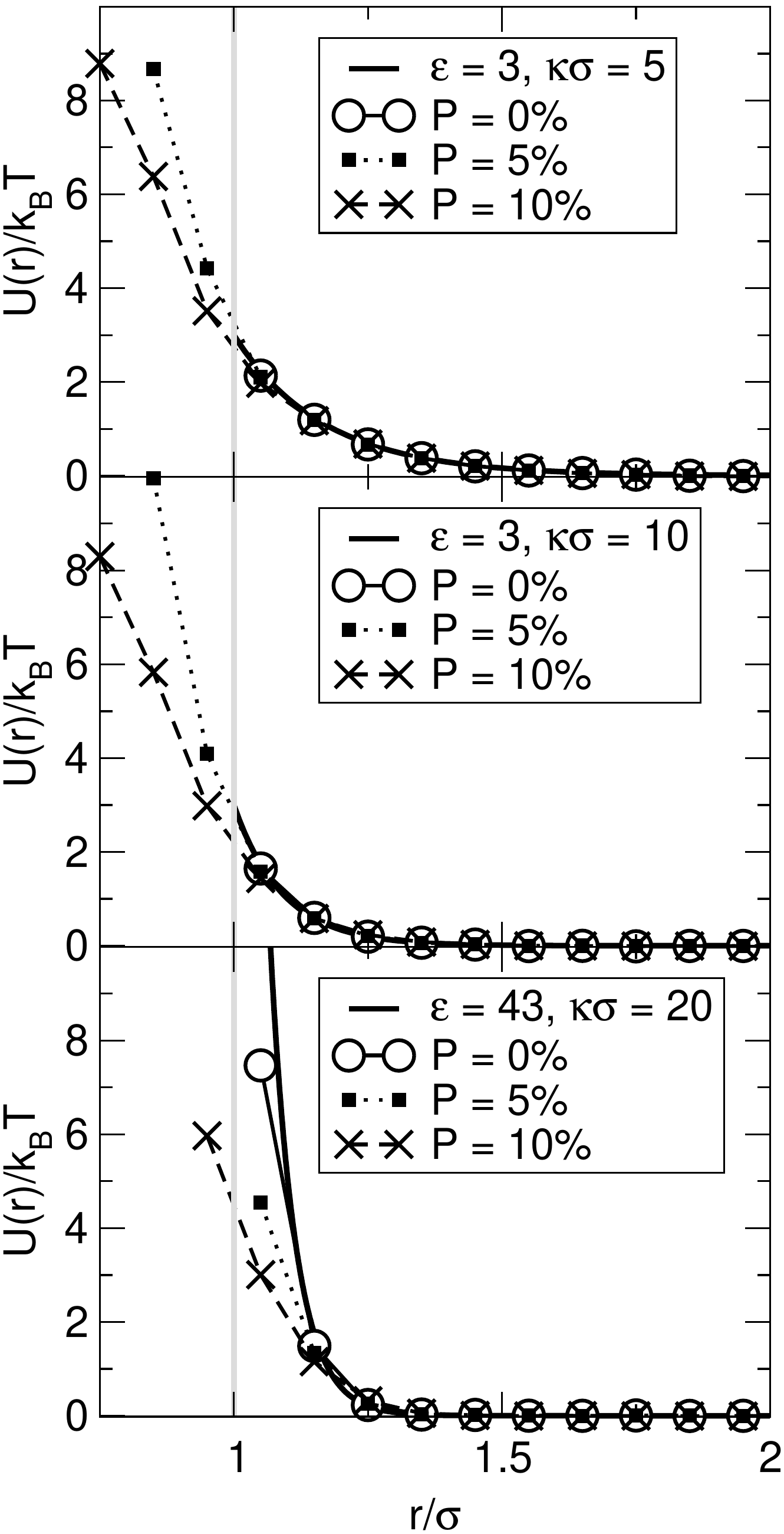} 
\caption{Comparison of extrapolated potentials with different levels of polydispersity. The grey line denotes the excluded volume region, which corresponds the minimum separation between two average sized particles.}
\label{figurePotentialPoly}
\end{figure}

From this we can conclude that as the range of the potential decreases, the effect of polydispersity increases.  This can be understood as arising from the fact that as the potential acts over shorter distances, the likelihood of close contact increases.  In a polydisperse system, where a subset of particles have diameters smaller than the average, we will observe a greater number of particles at small separations of $g(r, \phi)$ within the excluded volume region.  Thus, we can expect that relatively short-ranged potentials (or those that are effectively short-ranged) will show a stronger effect of polydispersity and our measurements of $\epsilon$ will be less accurate than $\kappa$. As a general rule, we should exclude any values of the derived potential that occur for $r/\sigma < 1$ and apply extra scrutiny to any values at small $r/\sigma$, especially in systems where the derived potential appears short-ranged.  Additionally, any potential estimation method that relies on $g(r,\phi)$ will suffer from this same issues regarding polydispersity and very short-ranged potentials.

\section{Experimental Results \label{resultsExp}}
\subsection{Analysis of Pure DOP System}

To test the validity of the potential derivation method in experiment, we first considered PMMA/PHSA particles in pure DOP.  We collected data for a range of $\phi$ = (0.005, 0.007, 0.008, 0.012, 0.018, 0.022, 0.042), as summarized in table \ref{particlesimagevolume1}. For the three lowest values of $\phi$ = (0.005, 0.007, 0.008), a large number of samples volumes were used for statistical purposes (see appendix regarding error scaling in $g(r)$).  Figure \ref{figureRDFpureDOP} plots the average $g(r,\phi)$ for $\phi$ = (0.005 - 0.022) where the bin shell size $\Delta r$ = 0.1$\mu$m (~10$\%$ of the particle diameter) to balance the competing constraints of signal to noise and spatial resolution of the potential.  At these charge and solvent conditions, we observed no particles in the inner most shells of $g(r,\phi)$.  This result is consistent with repulsive pair interactions, as reported for similar systems \cite{royall2003,royall2007}.  Error bars represent the standard error of the mean of $g(r,\phi)$.
\bigskip
\bigskip

\begin{figure}[ht]
\includegraphics[width=2.5in]{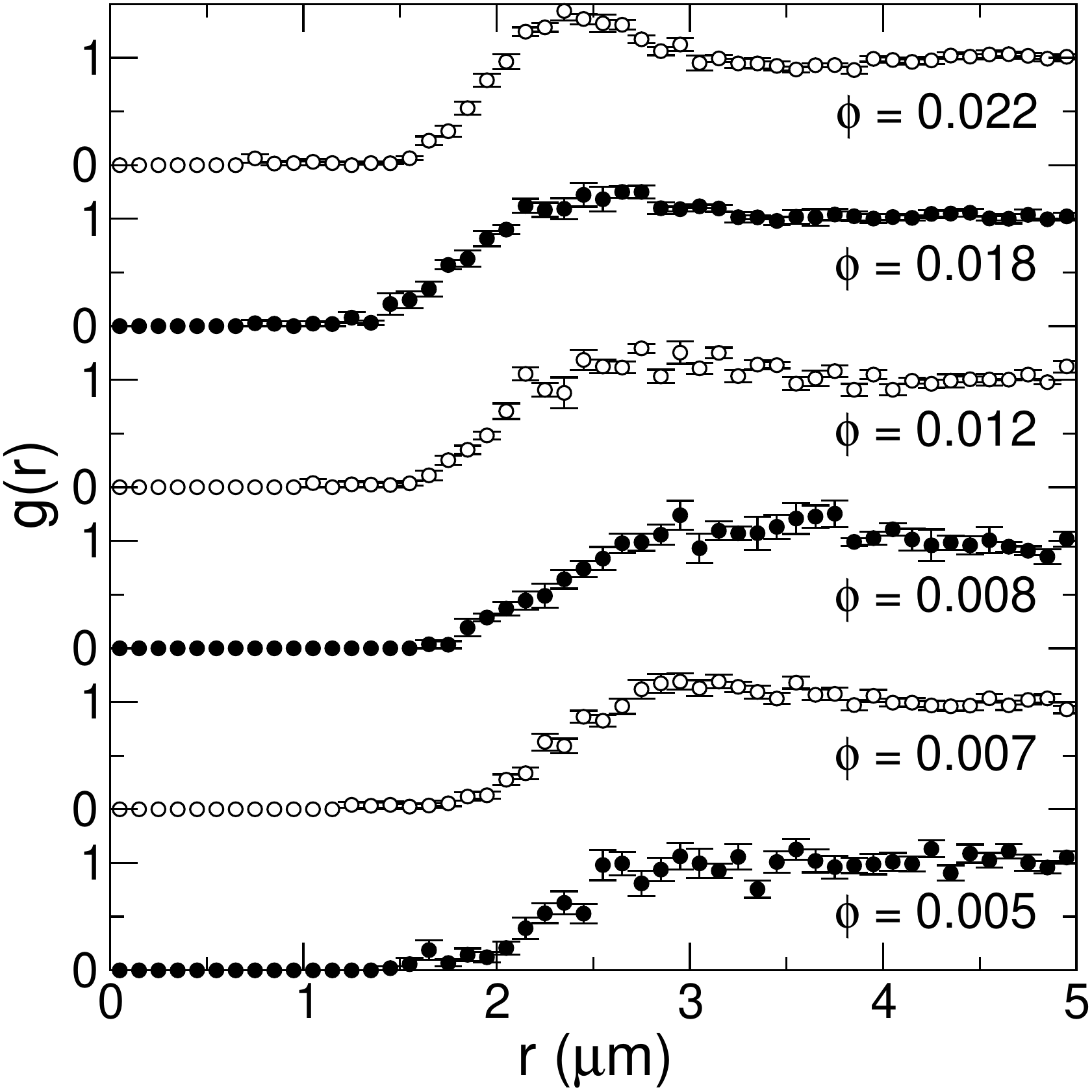} 
\caption{$g(r,\phi)$ for PMMA/PHSA particle system in pure DOP for $\phi$ = (0.005, 0.007, 0.008, 0.012, 0.018, 0.022) respectively from the bottom.}
\label{figureRDFpureDOP}
\end{figure}

\begin{table}[htdp]
\begin{center}
\begin{tabular}{c c c }
\quad \quad $\phi$\quad \quad \quad  & $N_{total}$ \quad & Total $\#$ of\\ & & image volumes \\
\hline
0.005 & 7216 & 36 \\
0.007 & 10911 & 26 \\
0.008 & 5772 & 12 \\
0.012 & 4686 & 6 \\
0.018 & 6842 & 6 \\
0.022 & 8279 & 6 \\
0.042 & 15997 & 6\\
\hline
\end{tabular}
\caption{Pure DOP solution. $N_{total}$ refers to the total number of particles from the multiple image volumes used in the calculation of $g(r, \phi)$.}
\end{center}
\label{particlesimagevolume1}
\end{table}

Figures \ref{potentialmeanforce}A-C plot $-ln[g(r,\phi)]$ (i.e. $W(r, \phi)$) as a function of $\phi$ for 0.005 $ < \phi < $ 0.042 for r = (1.75, 1.85, 1.95) $\mu$m for the pure DOP system.  As we saw in the simulations (see Figure \ref{figurePhiMaxvR}), there is a clear region that displays linear behavior; the crossover between ``dilute'' and ``concentrated'' appears at  $\sim$ 0.0125 which is nearly identical to the value of $\phi_{max, median} \sim$ 0.015 approximated from Figure \ref{figurePhiMaxScaling} for $\epsilon$ =27 and $\kappa \sigma$ =2.2 from table \ref{particlechar}.

\begin{figure}[ht]
\includegraphics[width=3.25in]{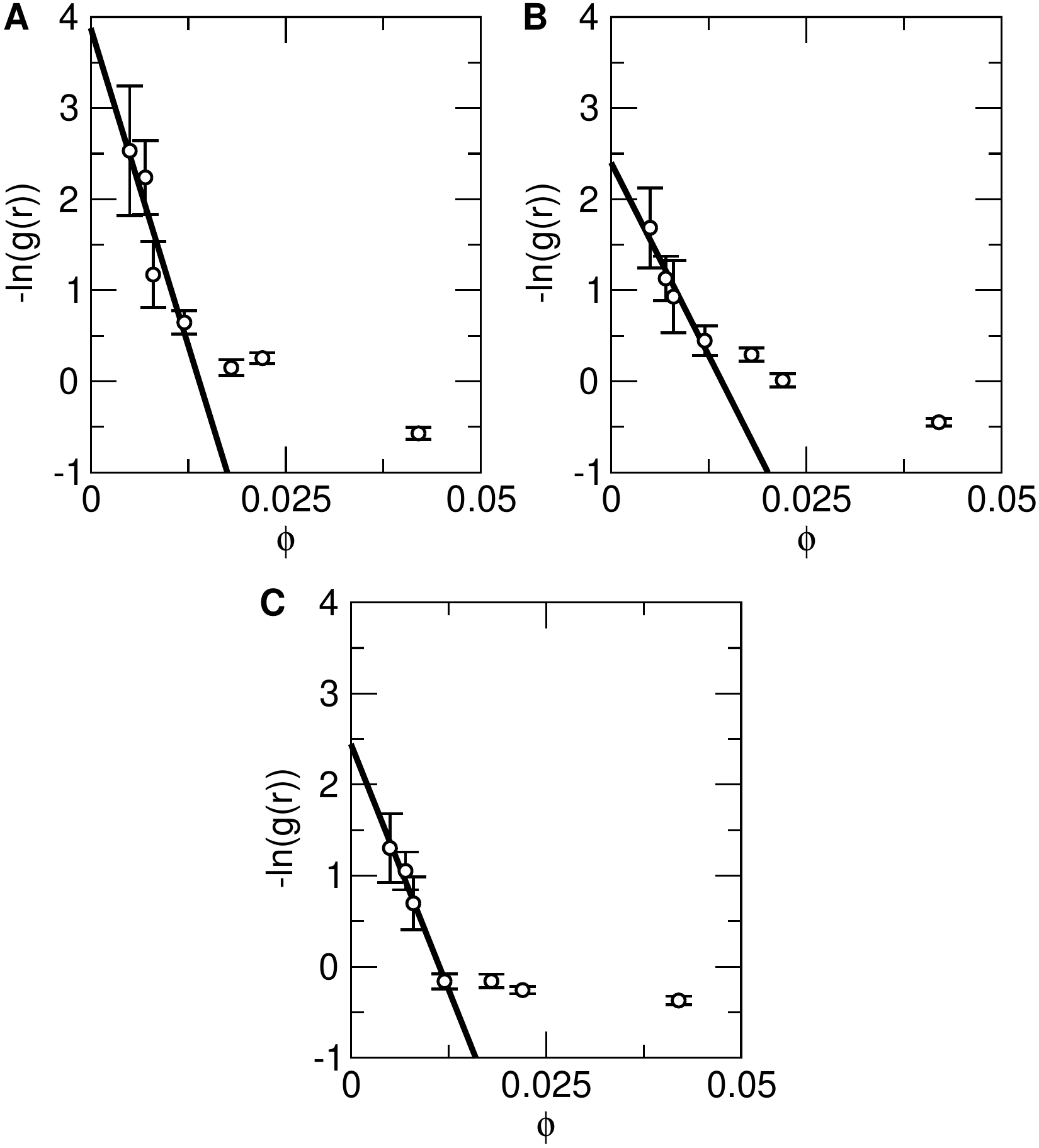} 
\caption{$W(r,\phi)$  vs. $\phi$ for  radial positions of (A) r = 1.75 $\mu$m, (B) r = 1.85 $\mu$m, and (C) r = 1.95 $\mu$m for the pure DOP system.  The linear, ``dilute'' regime appears for $\phi <$ 0.0125.}
\label{potentialmeanforce}
\end{figure}

\begin{figure}[ht]
\includegraphics[width=3.75in]{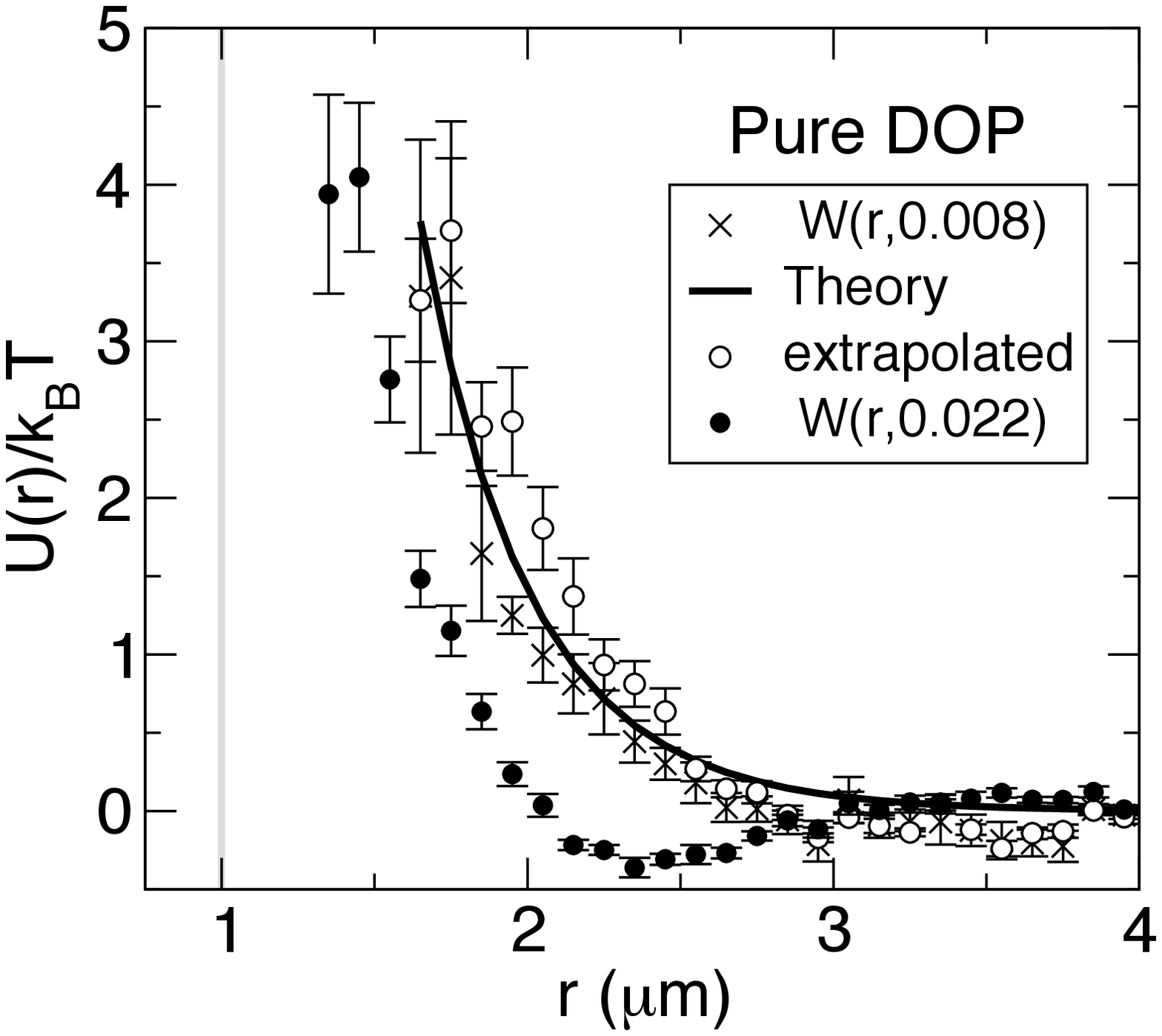} 
\caption{$U(r)$ for PMMA/PHSA particle system in  pure DOP. Screened Coulomb potential calculated from electrokinetics/fitting is shown as a solid line, the extrapolated potential is shown as open circles.  $W(r,0.022)$ is plotted as filled circles, and $W(r,0.008)$ is plotted with x symbols.  The gray line denotes the excluded volume region appearing at r $<$ 1 $\mu$m.}
\label{PotentialpureDOP}
\end{figure}

The pair potential calculated using the extrapolation method for PMMA/PHSA in the pure DOP solvent is plotted Figure \ref{PotentialpureDOP}.  $U(r)/k_BT$ is qualitatively consistent with long-range repulsive interactions, as reported for similar systems \cite{royall2003}.  Error bars plotted are standard error showing the data is precise.  The theoretical description of the potential (derived from electrokinetics and the one parameter fit of the unknown solvent conductivity) is also plotted in Figure \ref{PotentialpureDOP} as reported in table \ref{particlechar}.   These two potentials are in agreement; the average magnitude of the deviation between the extrapolated potential and the theoretical prediction is $M$ = 0.2.  Note that we expect the value of $M$ to be greater for the experimental system than simulation due to experimental noise; e.g. $M$= 0.15 when comparing the raw data to a best fit of the form of the Yukawa potential.  We additionally plot $W(r,0.022)$ and $W(r,0.008)$.  As we saw in our simulations, $W(r,\phi)$ calculated from even relatively low $\phi$ values may drastically underestimate the potential (see Figure  \ref{PotentialpureDOP}), and the only way to determine this would be to collect additional datasets. In Figure \ref{PotentialpureDOP} we plot $W(r,0.022)$ where $M$ = 0.65, noting that even this relatively low volume fraction provides a poor estimate of the potential. We also plot $W(r,0.008)$ which matches both the extrapolated and theory potential well, with $M$ = 0.18.

\subsection{Effect of TBAC Addition \label{resultsExpTBAC}}
To assess the extent to which the method can resolve differences in pair potential interactions that are relevant for self-assembly, we performed experiments in which the range of repulsive interactions are reduced by the addition of electrolyte.  We followed the same procedure previously outlined.  We collected data at $\phi$ = (0.012, 0.019, 0.024, 0.037, 0.049) for the 10$\mu$M TBAC system and at $\phi$ = (0.012, 0.018, 0.031, 0.039, 0.052) for the 2mM TBAC system.  For both TBAC concentrations, we observed that the onset of finite $g(r, \phi)$ is shifted to smaller radial distances (see appendix for more detail).  This shift qualitatively indicates that TBAC addition moderates repulsive interactions in the system.  This shift also affects the limit of linearity.  For example, for the two cases of added salt ( [TBAC] = 10 $\mu$M and 2 mM) we found that linearity was maintained out to $\phi \sim$ 0.05 which corresponded to the highest volume fraction studied. ÊThis increase in the limit of linearity is consistent with the simulations, for which $\phi_{max,median} \sim 0.1$ from Figure \ref{figurePhiMaxScaling}. ÊThat is, compared to the case of pure DOP, where linearity is found only forÊ~$\phi<$ 0.0125, these results coincide with an expected shift due to an increase in the screening of the repulsive interactions shown in simulation.

\begin{figure}[ht]
\includegraphics[width=3.25in]{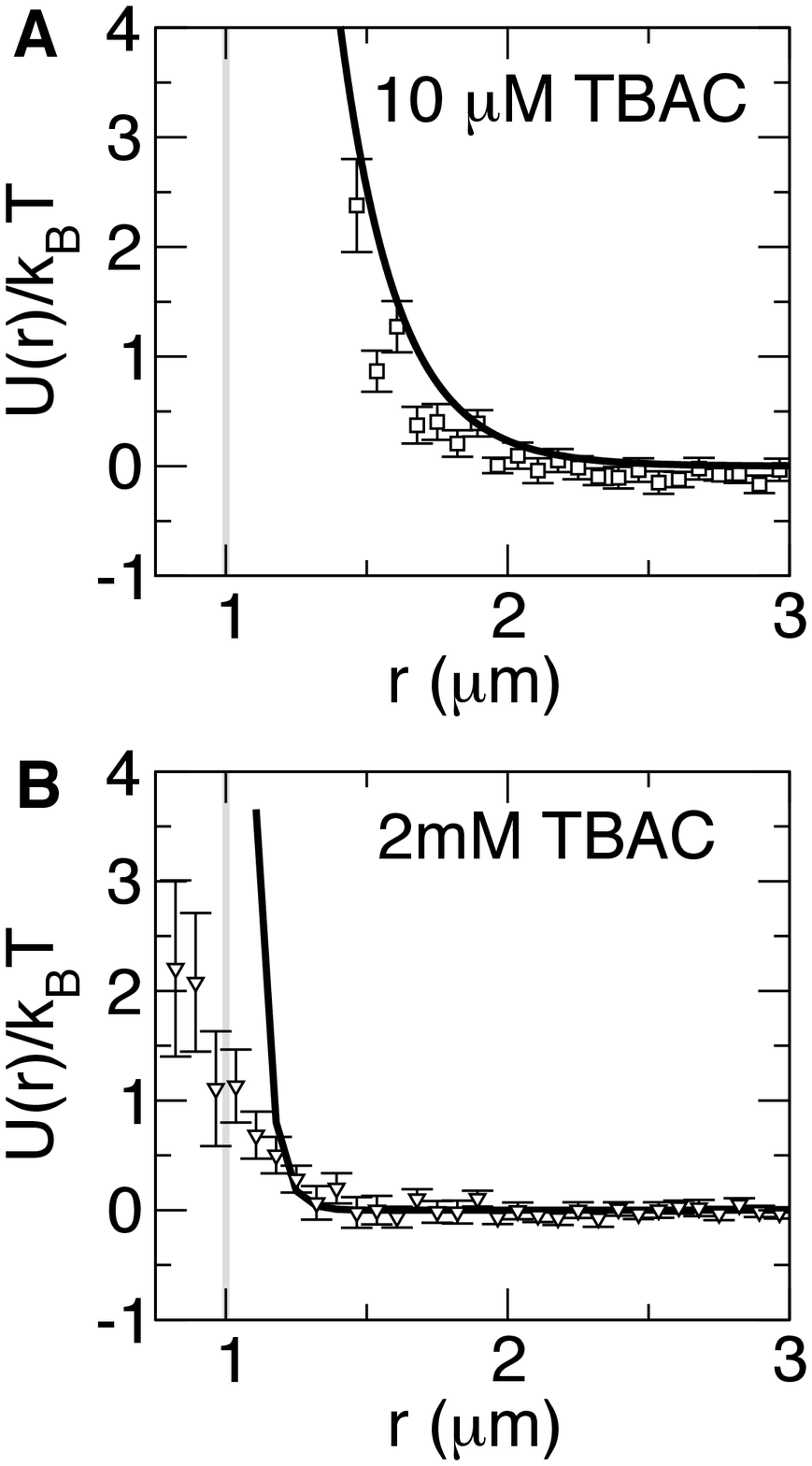} 
\caption{$U(r)$ for PMMA/PHSA particle system for  (A) 10$\mu$M TBAC, and (B) 2mM TBAC.  Screened Coulomb potentials calculated from electrokinetics/fitting are shown as solid lines, extrapolated potentials are shown as symbols.  The gray line denotes the excluded volume region appearing at r $<$ 1 $\mu$m.}
\label{Potentials}
\end{figure}

The extrapolated pair potentials for the systems with TBAC  are plotted in Figure \ref{Potentials}.  Visually, we find good agreement between the theoretical potential calculated from electrokinetics for 10$\mu$M system where $M$ = 0.35, as shown in Figure \ref{Potentials}A .  For [TBAC] = 2 mM, we find that the extrapolated potential does not provide as close a match as the pure DOP and 10$\mu$M systems when compared to electrokinetic-based potential; for [TBAC] = 2 mM, $M$ = 0.84.   The 2mM TBAC system is nearly consistent with simple excluded volume interaction; the transition to the repulsive portion of the potential is abrupt and very close to the measured diameter of the colloid, with a Debye length of 49 nm.  Thus, screened electrostatic interactions extend no further than about 5$\%$ of the particle diameter, not too different from the spread in the particle size distribution due to polydispersity which was calculated to be 4$\%$.  The effect of polydispersity on this 2mM system is consistent with what was observed in simulation, as discussed previously in Figure \ref{figurePotentialPoly}; as the range of the potential decreases, polydispersity plays a stronger role and the observed $\epsilon$ value decreases. However, if we truncate the extrapolated potential at r $\sim 1.25 \mu m$, ignoring any interactions within the excluded volume region and close to the surface of an average particle, we arrive at a more satisfactory description of $U(r)$. We should also note that, as seen in the simulations, we do not resolve $U(r)/k_BT >$ 10 for any of the systems studied; however, even given this constraint we can still easily resolve the shape of the potential. 

\section{Conclusions \label{conclusions}}
We have presented a method to calculate the pair potential, $U(r)$, between colloidal particles from microscopy.  This method relies on extrapolating the potential of mean force, $W(r,\phi)$, to $\phi$ = 0.  We have shown using MC simulation that this method produces near perfect results for ideal monodisperse colloids.  We have used simulation to explore the impact of surface charge and screening length on this method, providing general guidelines for the use of this method over a wide range of parameters.  We have also demonstrated that low levels of polydispersity have only a small impact on the accuracy of the calculation of $U(r)$ for longer-ranged potentials, however the effect is more pronounced as the range of the potential decreases.  Further, we have applied this method to experimental colloidal particles using the guidelines established from simulation.  We found close agreement of the results of the extrapolation method with theoretical screened Coulombic potentials.  As was noted in simulation, the effect of polydispersity becomes stronger for short ranged potentials where the range of the potential is approximately equal to the spread in particle diameter.  Moreover, we see that this methodology is well suited to determine the potential for a range of systems, including those that are refractive index matched, particularly for systems where the screening length is on the order of the particle diameter and many body effects are minimal.  We anticipate that this methodology can also be used to quantify $U(r)/k_bT$ for attractive systems, assuming the attraction is not so strong as to induce gelation. Additionally, this method can be used in concert with electrokinetic measurements as a means to double-check their validity or used in their place when it is not possible to measure key parameters (e.g. conductivity).

\section{Acknowledgements}
We thank Lilian Hsiao and Emcee Electronics, Inc. (Venice, Florida) for measurement of the conductivity of the solvent of 10$\mu$M TBAC in dioctyl phthalate.

%\bibliography{potentials.bib}
%merlin.mbs 2010-03-15 4.21a (PWD, AO, DPC)

\section{Appendix}
\subsection{Addition of TBAC to the DOP solvent}
In section \ref{resultsExpTBAC} we explored the addition of TBAC to the DOP solvent.  The total number of image volumes/particles collected are summarized in table \ref{tbacsup10} for the 10 $\mu$M solution and \ref{tbacsup2} for the 2mM solution.  The $g(r,\phi)$ data used to calculate the extrapolated $U(r)/k_bT$ (extrapolations were previously shown in Figure \ref{PotentialpureDOP}) are plotted for in Figure \ref{figureGrTBAC}.

\bigskip

\begin{table}[htdp]
\begin{center}
\begin{tabular}{c c c }
\quad \quad $\phi$\quad \quad \quad  & $N_{total}$ \quad & Total $\#$ of\\ & & image volumes \\
\hline
0.012 & 6646 & 10\\
0.019 & 8331 & 8\\
0.024 & 10434 & 8\\
0.037 & 12183 & 8\\
0.049 & 16116 & 6\\
\hline
\end{tabular}
\caption{[TBAC] = 10 $\mu$M.}
\end{center}
\label{tbacsup10}
\end{table}

\begin{table}[htdp]
\begin{center}
\begin{tabular}{c c c }
\quad \quad $\phi$\quad \quad \quad  & $N_{total}$ \quad & Total $\#$ of\\ & & image volumes \\
\hline
0.012 & 10568 & 14\\
0.019 & 13508 & 12\\
0.031 & 10196 & 6\\
0.037 & 12453 & 6\\
0.049 & 15129 & 6\\
\hline
\end{tabular}
\caption{[TBAC] = 2 mM.}
\end{center}
\label{tbacsup2}
\end{table}

\bigskip

\begin{figure}[ht]
\includegraphics[width=1.65in]{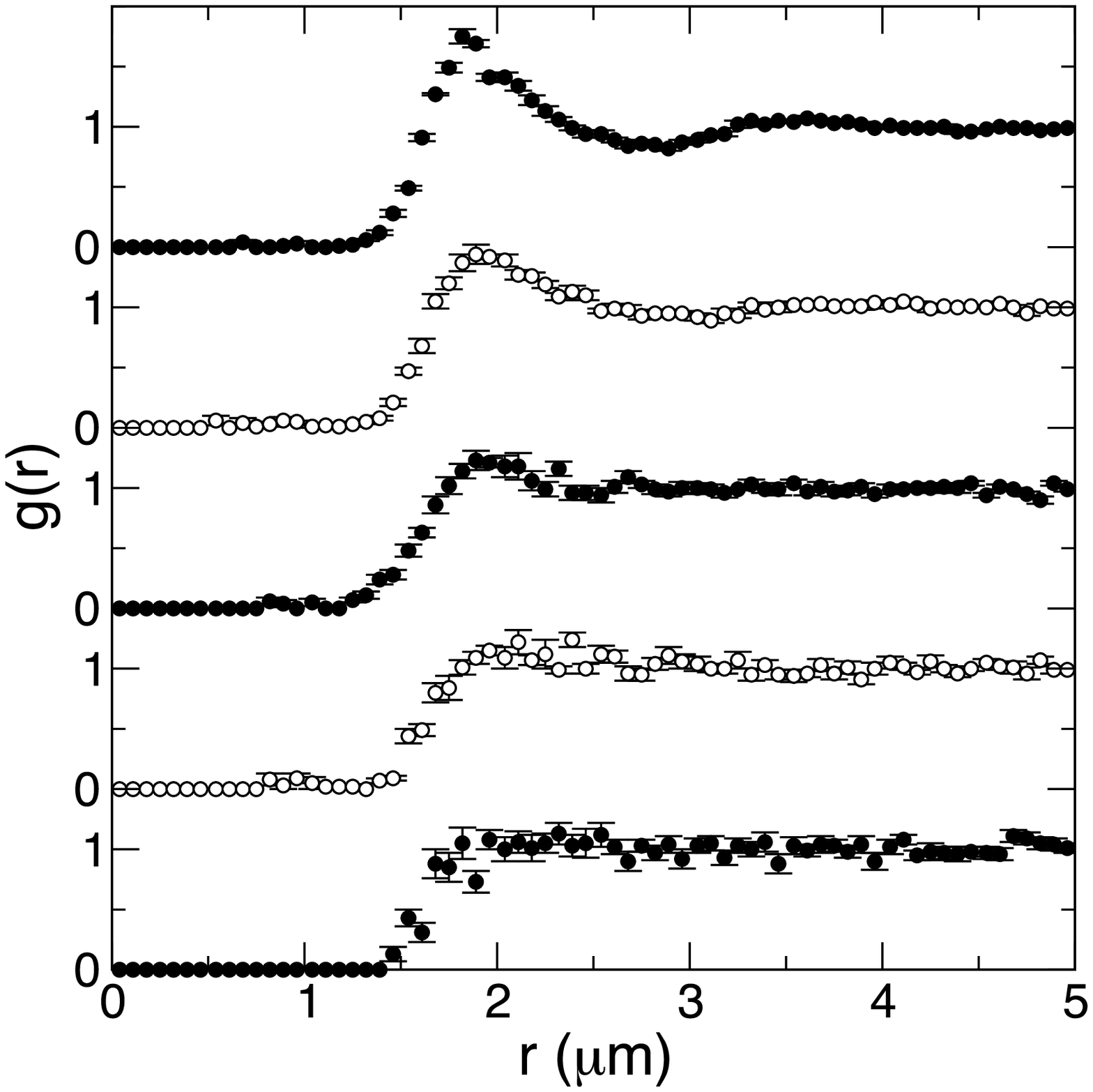} 
\includegraphics[width=1.65in]{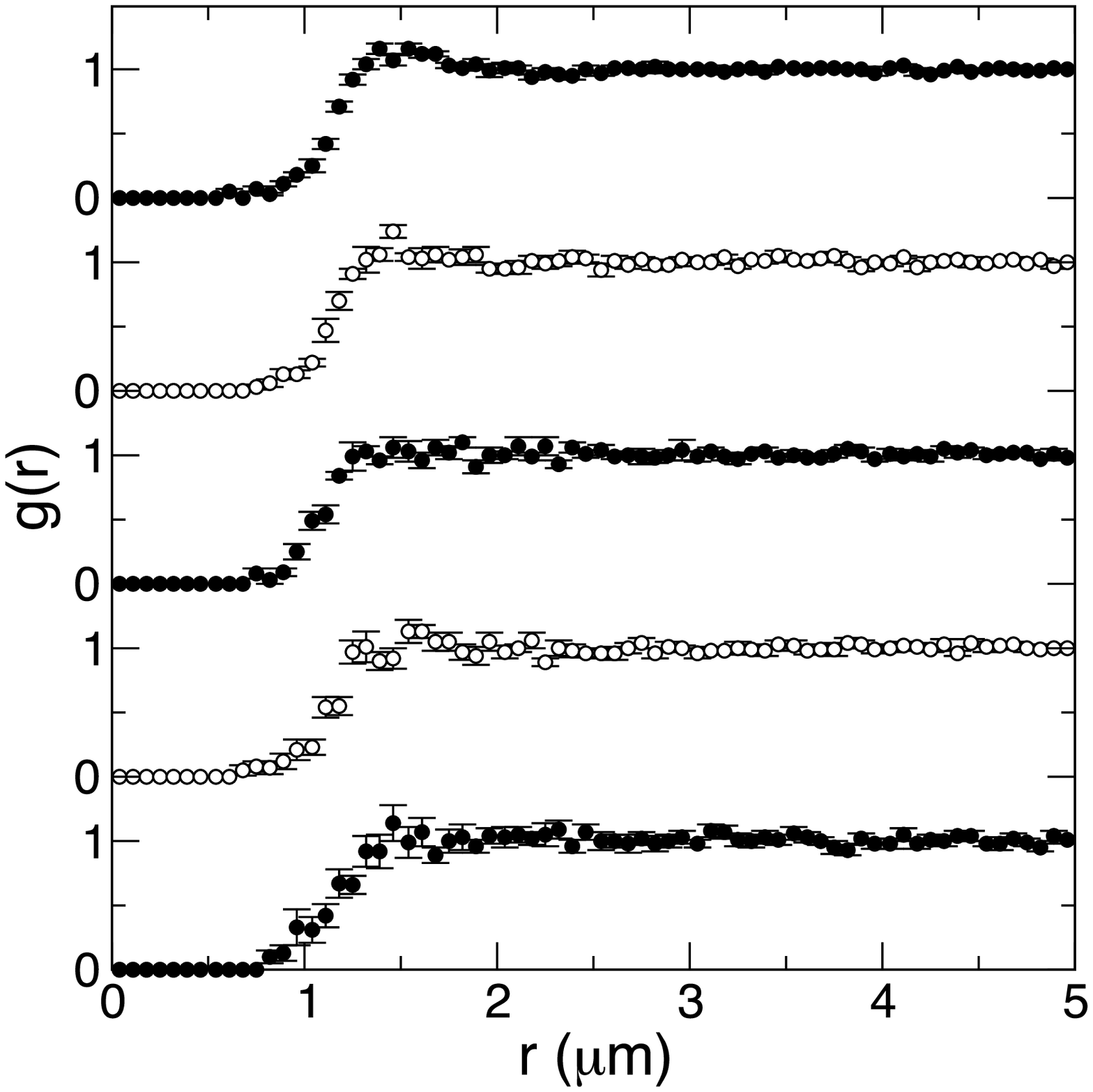} 

\caption{$g(r,\phi)$ plots for [TBAC]= 10 $\mu$M (left) and [TBAC] = 2 mM (right).  Data corresponds to $\phi$ = (0.012,0.019,0.024, 0.037, 0.049)  [TBAC] = 10 $\mu$M (left)and $\phi$ = (0.012,0.019,0.031, 0.037, 0.049) for [TBAC] = 2 mM (right) respectively from the bottom.}
\label{figureGrTBAC}
\end{figure}

\subsection{Error scaling in the radial distribution function}
Essential to accurately determining the pair potential is generating $g(r,\phi)$ with sufficient accuracy.   $g(r,\phi)$ is a measure of the probability of finding a particle at a distance \textit{r} away from a given reference particle.   This distribution function is determined by calculating the distance between all particles pairs and binning them into a histogram;  the histogram is then normalized with respect to an ideal gas, where particle histograms are completely uncorrelated.  Due to its construction,  $g(r,\phi)$ is a spherically averaged measure; therefore, we average over more particle pairs as we consider larger particle separations.  In other words, our statistics increase as \textit{r} increases.  This may be problematic since the pair potential we wish to determine acts the strongest at smaller separations where we have the least statistics.  Additionally, since we must compute $g(r,\phi)$ at low values of $\phi$, it is unlikely to have many particles interacting at small values of \textit{r} if particles are well distributed in the system (i.e. particles do not aggregate); unlike in a dense system, a single configuration snapshot will not be sufficient to accurately generate $g(r, \phi)$.  As such, to determine $U(r)$ accurately, we must establish when we have appropriately gathered sufficient statistics to be confident in $g(r,\phi)$.

\begin{figure}[ht]
\includegraphics[width=3.25in]{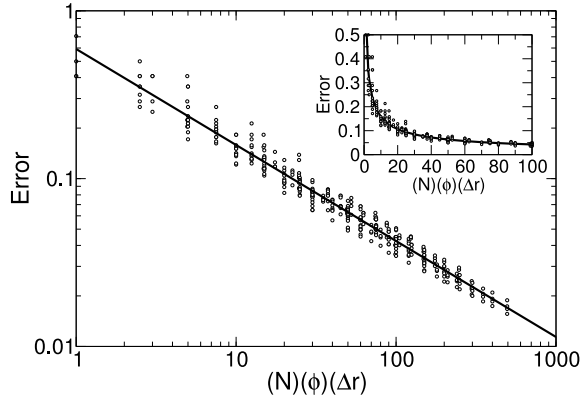} 
\caption{Error scaling in $g(r,\phi)$ for a combination of Yukawa systems with $\epsilon=3$ and $\kappa = (3.0, 1.5, 0.5, 0.25)$, over a range of volume fractions, $\phi = (0.001, 0.005, 0.01, 0.02$, and $\Delta r = (0.05, 0.15, 0.25)$.}
\label{figureErrorScaling}
\end{figure}

We considered the error in $g(r,\phi)$ at a given separation of $r$ to be proportional to $1/\sqrt{N_{bin}}$, where $N_{bin}$ is the number of particles in the bin.  By this construction, our maximum error is $1/\sqrt{2} = 0.707$ and our minimum error will approach 0.  There are three main factors that impact this value: the error decreases as we increase the total number of particles considered, $N_{total}$ (i.e. the sum of all particles in all the samples used to construct $g(r,\phi)$); the error decreases as we increase our volume fraction, $\phi$; and the error decreases as we increase the width of the bin, $\Delta r$.  In Figure \ref{figureErrorScaling} we plot $1/\sqrt{N_{bin}}$ vs. $(N_{total})(\phi)(\Delta r)$ for a combination of Yukawa systems;  here we have chosen to plot the data for a bin close to the surface of the particle, specifically $r/\sigma$ = 1.125, as this should be characteristic of the maximum error in the system.  We simulated four different potentials; we fixed $\epsilon=3$ and varied the inverse screening length, specifically $\kappa$ = (3.0, 1.5, 0.5, 0.25).  For each potential we performed simulations over a range of volume fractions, $\phi$ = (0.001, 0.005, 0.01, 0.02) and then analyzed each of these simulations using three bin sizes, $\Delta r$ = (0.05, 0.15, 0.25).  We plotted the data from these four potentials in Figure \ref{figureErrorScaling} finding that the error roughly scales as a power law with exponent -0.57. The inset of Figure \ref{figureErrorScaling} plots the data on a standard axis; it is clear from this plot that our error is most rapidly decreasing in the range of  $1 < (N_{total})(\phi)(\Delta r) < 10$.  At values of $(N_{total})(\phi)(\Delta r) > 10$, our error decreases very slowly.  Thus for the best statistics in the $g(r,\phi)$, a rough guideline should be to perform enough samples such that $(N_{total})(\phi)(\Delta r)>10$.   For our simulations $\Delta r$ = 0.1 and $(N_{total}$ = 1000000, thus, even for our lowest volume fraction of $\phi$= 0.001, this condition is met.

Similarly, Figure \ref{ErrorPotential} summarizes the relationship between the precision of the measured $g(r,\phi)$ and properties of the experiments, including the number of particles, the volume fraction of the specimen, and the bin size of each point in  $g(r,\phi)$.  We see that the error in $g(r,\phi)$  is a weak function of the combination $(N_{total})(\phi)(\Delta r)$, scaling as power law with exponent -0.79.  Figure \ref{ErrorPotential} can be applied to determine experimental conditions for pair potential characterization.  For example, for a relative error of $\sim$10$\%$ in the radial distribution, Figure \ref{ErrorPotential} requires $(N_{total})(\phi)(\Delta r) \sim$ 7.  For a volume fraction 0.01 with 1 $\mu$m particles in which $\Delta r$ = 0.1 $\mu$m, this relative error would require  on the order of 7000 particles.  

\begin{figure}[ht]
\includegraphics[width=3.25in]{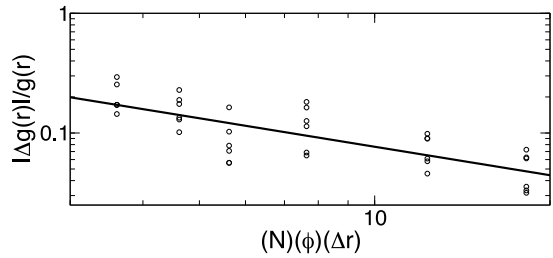} 
\caption{Error scaling in $g(r,\phi)$ as a function of number of particles, volume fraction, and bin size for pure DOP, no salt case.  Data taken from four different radial positions, r = (2.05, 2.15, 2.35, 2.55, 2.75, 3.05) $\mu$m.  Line drawn is power law fit with exponent -0.79.}
\label{ErrorPotential}
\end{figure}

\end{document}